\documentclass[10pt, aps, prb, twocolumn, superscriptaddress, showpacs]{revtex4-1}

\usepackage{here}
\usepackage{amsmath}
\usepackage{graphicx}
\usepackage{braket}
\usepackage{lettrine}
\usepackage{color}
\usepackage{times}
\usepackage{mathtools}
\usepackage{tabularx}

\begin{document}

\title{Gigantic intrinsic orbital Hall effects in weakly spin-orbit coupled metals}

\author{Daegeun Jo}
\affiliation{Department of Physics, Pohang University of Science and Technology, Pohang 37673, Korea}

\author{Dongwook Go}
\email{godw2718@postech.ac.kr}
\affiliation{Department of Physics, Pohang University of Science and Technology, Pohang 37673, Korea}

\author{Hyun-Woo Lee}
\email{hwl@postech.ac.kr}
\affiliation{Department of Physics, Pohang University of Science and Technology, Pohang 37673, Korea}


\begin{abstract}
A recent paper [Go {\it et al.}, Phys. Rev. Lett. {\bf 121}, 086602 (2018)] proposed that the intrinsic orbital Hall effect (OHE) can emerge from momentum-space orbital texture in centrosymmetric materials. In searching for real materials with strong OHE, we investigate the intrinsic OHE in metals with small spin-orbit coupling (SOC) in face-centered cubic and body-centered cubic structures (Li, Al, V, Cr, Mn, Ni, and Cu). We find that orbital Hall conductivities (OHCs) in these materials are gigantic $\sim 10^3-10^4\ (\hbar/e)(\Omega\cdot\mathrm{cm})^{-1}$, which are comparable or larger than spin Hall conductivity (SHC) of Pt. Although SHCs in these materials are smaller than OHCs due to small SOC, we found that SHCs are still sizable and the spin Hall angles may be of the order of 0.1. We discuss implications on recent spin-charge interconversion experiments on materials having small SOC.
\end{abstract}

\maketitle

\section{Introduction}\label{sec:introduction}

The spin Hall effect (SHE) refers to a phenomenon where an external electric field drives a transverse spin current, leading to boundary accumulation of the spin.\cite{dyakonov1971, hirsch1999, engel2007, schliemann2006, sinova2015} In combination with its reciprocal phenomenon called the inverse SHE, it allows for electrical generation and detection of the spin, which is essential for spintronic device applications.\cite{saitoh2006, kimura2007, niimi2015} In heavy elements, such as Pt and Ta, the SHE can be used to inject the spin current to an adjacent ferromagnetic layer and control the magnetization direction electrically. \cite{liu2012prl, liu2012science} The large SHE in these materials is attributed to an intrinsic mechanism{\cite{tanaka2008, kontani2009, guo2008, freimuth2010, morota2011, sagasta2016}} which does not rely on impurity scatterings.

However, theoretical understanding of the intrinsic SHE is rather limited. Sinova \emph{et al.} described the spin-texture-based mechanism of the intrinsic SHE.\cite{sinova2004} Unfortunately, it is applicable only to noncentrosymmetric materials, since the spin texture calls for inversion symmetry breaking. Hence, there have been efforts to explain the intrinsic SHE in centrosymmetric materials. Murakami \emph{et al.} showed that hole-doped GaAs exhibits intrinsic SHE from the Luttinger model.\cite{murakami}  Nonetheless, it cannot explain the sign change of the spin Hall conductivity (SHC) observed in first-principles calculation.\cite{yao2005}  Moreover, it is not straightforward to generalize its mechanism to other centrosymmetric materials. On the other hand, Tanaka and Kontani \emph{et al.} performed systematic study on $4d$ and $5d$ transition metals from the tight-binding (TB) model and succeeded in explaining a Hund's rule type behavior of the SHC.\cite{tanaka2008, kontani2009} They attributed its mechanism to the orbital Hall effect (OHE) originating from the orbital Aharonov-Bohm effect.\cite{tanaka2008, kontani2009} However, its relation to the electronic structure is not yet clear, i.e. how to engineer the band structure to enhance the OHE and SHE. 
Also, since the crystal field destroys orbital angular momentum, which is called the orbital quenching,\cite{kittel2004} it is not certain whether the OHE would lead to any observable effect such as boundary accumulation.

Despite numerous studies from both model\cite{murakami, murakami2004prl, tanaka2008, kontani2009} and first-principles\cite{guo2008, freimuth2010, yao2005} calculations, there has been yet no simple and general quantum mechanical model of the intrinsic SHE for bulk centrosymmetric materials until very recently. With the above issues in mind, we recently showed that momentum-space orbital texture can give rise to the intrinsic OHE.\cite{go} The OHE in combination with spin-orbit coupling (SOC) generate the SHE as well since SOC correlates spin and orbital degree of freedom,\cite{go} which was already pointed out in Refs. \onlinecite{tanaka2008, kontani2009, kontani2008}. Here, the orbital texture means the varation of the orbital characters along with the crystal momentum $\mathbf{k}$. We emphasize that the orbital texture can exist quite generally regardless of whether the inversion symmetry is present or broken, while the spin texture requires the breaking of the mirror symmetry. The orbital-texture-based mechanism not only enables microscopic understanding of the intrinsic OHE and SHE, but also provides alternative understanding of the orbital Aharonov-Bohm effect picture in Refs. \onlinecite{tanaka2008, kontani2009, kontani2008}. More importantly, the orbital-texture-based mechanism can naturally explain the stability of the intrinsic OHE against the orbital quenching.\cite{go} Now that a general principle of the intrinsic OHE is understood, it is desirable to find real materials exhibiting large orbital Hall conductivity (OHC) to provide a guideline for an experimental observation.

\begin{figure}[t]
	\includegraphics[angle=0, width=0.47\textwidth]{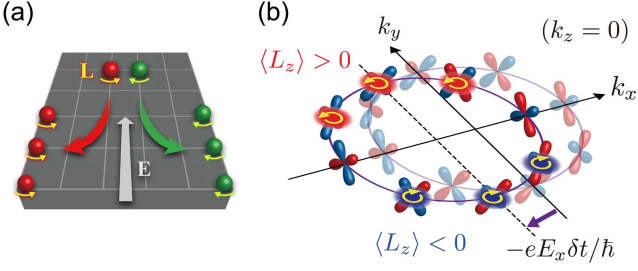}
	\caption{(a) Schematic illustration of the OHE. The angular momentum $\mathbf{L}$ is defined from localized orbitals around the atom at each lattice. In the presence of an external electric field $\mathbf{E}$, electrons with opposite $\mathbf{L}$ deflect in the clockwise (red arrow) or anticlockwise (green arrow) direction. (b) Illustration of the mechanism of the intrinsic OHE based on the momentum-space orbital texture. Shifting of the Fermi surface induces dynamics of the orbitals, generating finite $\left\langle L_z \right\rangle$. Since the dynamics of the orbitals occurs in the opposite way for positive and negative $k_y$'s, induced $\left\langle L_z \right\rangle$ is positive(negative) for a state with positive(negative) $k_y$, which results in the OHE.} 
	\label{fig:ohe}
\end{figure}

Before proceeding further, we define what we mean by the OHE in this paper. We refer to the OHE as a flow of the orbital angular momentum defined around atoms at each lattice to a perpendicular direction of an external electric field [Fig.~\ref{fig:ohe}(a)]. Note that our definition of the orbital angular momentum excludes a contribution from the itinerant circulation.\cite{thonhauser2005, xiao2005} Instead, we consider localized orbitals around atomic core, i.e. $\ket{d_{zx}}$ and $\ket{d_{yz}}$ states. Although these cubic harmonic states carry no orbital angular momentum \emph{in equilibrium}, an external electric field induces hybridization of different orbitals, such as $\ket{d_{zx}}\pm i\ket{d_{yz}}$, leading to finite orbital angular momentum in \emph{nonequilibrium}. Then we find that the states with the opposite signs of orbital angular momentum travel to the opposite directions as schematically shown in Fig.~1(a).

In this paper, we first demonstrate the orbital-texture-based mechanism \cite{go} for the example of Pt, one of the $sd$ metals. In Pt, the orbital texture from the $d$ orbitals is crucial for the OHE [Fig.~\ref{fig:ohe}(b)]. By tuning SOC strength of Pt, a correlation between the OHE and SHE is discussed. We emphasize that large SOC does not always guarantee large SHC because it may decrease the OHC. We next consider materials with small SOC (Li, Al, V, Cr, Mn, Ni, and Cu). We find that the OHCs in these materials are in general gigantic $\sim 10^3-10^4\ (\hbar/e)(\Omega\cdot\mathrm{cm})^{-1}$, which are comparable or larger than the SHC of Pt $\sim 2000 \ ( \hbar/ e ) ( \Omega \cdot \mathrm{cm} )^{-1} $. We emphasize that this is in contrast to a common expectation that the orbital degree of freedom is quenched in solids unless the SOC is substantial. Furthermore, although these materials have weak SOC, we find that their SHCs are as large as $\sim 10^2\ (\hbar/e)(\Omega\cdot\mathrm{cm})^{-1}$, which is not negligible thus experimentally measurable. Such gigantic OHCs and sizable SHCs in weak SOC materials may appear counterintuitive but are natural in view of Ref. \onlinecite{go}: the OHE is stable even without SOC and the SHE is converted from the OHE by SOC, that is to say, the OHE is more fundamental than the SHE. Thus if the OHCs are gigantic, the converted SHCs can be sizable even when SOCs are weak. Recent experiments have found that spin Hall angles (SHAs) in weak SOC materials, such as V,\cite{wang2017} Cr,\cite{qu2015} Ni,\cite{du2014} Py,\cite{miao2013} and $\textup{CuO}_x$\cite{an2016, gao2018} can be substantial. We discuss possible explanations of such large SHAs in these materials based on our calculation.

This paper is organized as follows. In Sec.~\ref{sec:methods}, the methods employed in this work are explained. In Sec.~\ref{sec:mechanism}, we demonstrate how the orbital-texture-based mechanism works for $d$ orbital systems. We consider Pt as an example in detail. In Sec.~\ref{sec:light}, we present the numerical results of the OHC and SHC for materials with small SOC: Al and Li as examples of the $sp$ metals and V, Cr, Mn, Ni, and Cu as examples of the $sd$ metals. In Sec.~\ref{sec:discussion}, we discuss the result from our numerical calculation and its implications on recent experiments.\cite{qu2015, du2014, miao2013, an2016, gao2018} Finally, Sec.~\ref{sec:summary} summarizes the paper.

\section{Methods}\label{sec:methods}

\subsection{Tight-binding model}\label{subsec:tb}

\begin{figure}[t]
	\includegraphics[angle=0, width=0.47\textwidth]{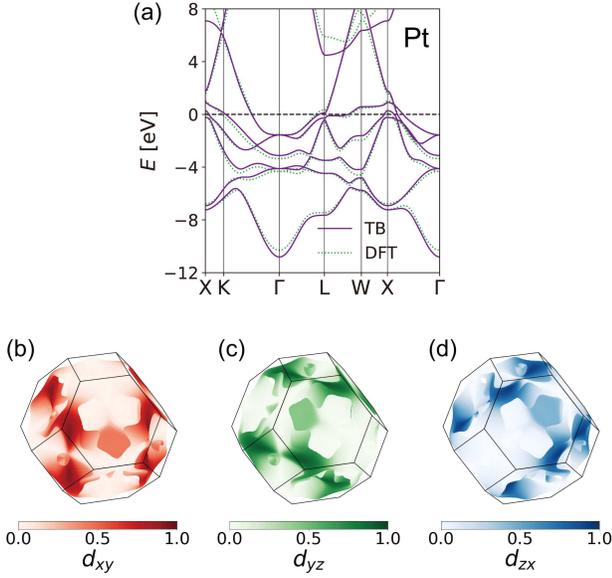}
	\caption{(a) Band structure of fcc Pt obtained from the TB and DFT calculations. Probability weightings of (b) $d_{xy}$, (c) $d_{yz}$, and (d) $d_{zx}$ orbital characters shown on top of the Fermi surface of Pt.} 
	\label{fig:orbital_character}
\end{figure}

We employ a TB model description to calculate energy bands and corresponding electronic wave functions of solids. Using atomic orbital basis in cubic harmonics, such as $s$, $p_{x}$, $p_{y}$, $p_{z}$, $d_{z^2}$, $d_{x^{2}-y^{2}}$, $d_{xy}$, $d_{yz}$, $d_{zx}$, we construct TB Hamiltonians from the Slater-Koster method.~\cite{slater1954} Parameters used in this calculation are taken from Ref.~\onlinecite{handbook}. 
We approximate SOC by 
\begin{equation}\label{eq:soc2}
	H_{\mathrm{soc}} = \sum_{l}  \frac{2\lambda_{l}}{\hbar^{2}} \mathbf{L}^{(l)}\cdot \mathbf{S},
\end{equation}
since SOC is most dominant near the atomic nuclei. \cite{bihlmayer} Here, $\lambda_{l}$ is the magnitude of the SOC from the $l$ orbitals ($l=s,p,d$), and $\mathbf{L}^{(l)}$ is the orbital angular momentum operator of $l$ orbitals near nuclei, which is defined with respect to atomic centers. For example, a matrix representation of $\mathbf{L}^{(p)} = (L_x^{(p)}, L_y^{(p)}, L_z^{(p)})$ is written as 
\begin{subequations}\label{orbital}
\begin{align}
	{L}_{x}^{(p)} = \hbar
	\begin{pmatrix} 
		0 & 0 & 0 \\
		0 & 0 & -i \\
		0 & i & 0 \\
	\end{pmatrix},  \\
	{L}_{y}^{(p)} = \hbar
	\begin{pmatrix} 
		0 & 0 & i \\
		0 & 0 & 0 \\
		-i &0 & 0 \\
	\end{pmatrix},  \\
	{L}_{z}^{(p)} = \hbar
	\begin{pmatrix} 
		0 & -i & 0 \\
		i & 0 & 0 \\
		0 &0 & 0 \\
	\end{pmatrix},
\end{align}
\end{subequations}
with basis states $\ket{p_x}$, $\ket{p_y}$, $\ket{p_z}$. Similarly, a matrix representation for $\mathbf{L}^{(d)}=(L_x^{(d)}, L_y^{(d)},L_z^{(d)})$ is
\begin{subequations}\label{orbital2}
	\begin{align}
	L_x^{(d)} &= \hbar
	\begin{pmatrix}
		0 & 0 & 0 & \sqrt{3}i & 0 \\
		0 & 0 & 0 & i & 0 \\
		0 & 0 & 0 & 0 & -i \\
		-\sqrt{3}i & -i & 0 & 0 & 0 \\
		0 & 0 & i & 0 & 0 
	\end{pmatrix}, \\
	L_y^{(d)} &= \hbar
	\begin{pmatrix}
		0 & 0 & 0 & 0 & -\sqrt{3}i \\
		0 & 0 & 0 & 0 & i \\
		0 & 0 & 0 & i & 0 \\
		0 & 0 & -i & 0 & 0 \\
		\sqrt{3}i & -i & 0 & 0 & 0
	\end{pmatrix},	\\
	L_z^{(d)} &= \hbar
	\begin{pmatrix}
		0 & 0 & 0 & 0 & 0 \\
		0 & 0 & -2i & 0 & 0 \\
		0 & 2i & 0 & 0 & 0 \\
		0 & 0 & 0 & 0 & i \\
		0 & 0 & 0 & -i & 0
	\end{pmatrix},
\end{align}
\end{subequations}
with basis states $\ket{d_{z^2}}$, $\ket{d_{x^2-y^2}}$, $\ket{d_{xy}}$, $\ket{d_{yz}}$, $\ket{d_{zx}}$. Trivially, $\mathbf{L}^{(s)}=0$.
In order to check the accuracy of the TB model, we compare the band structure obtained from the TB model and the density functional theory (DFT) calculations, and find a good agreement near the Fermi energy. Figure \ref{fig:orbital_character}(a) shows the comparison for face-centered cubic (fcc) Pt, for example. For further details of the calculation, see Appendix.

\subsection{Kubo formula}\label{subsec:kubo}

In order to calculate OHC ($\sigma_\mathrm{OH}$) and SHC ($\sigma_\mathrm{SH}$), we employ the Kubo formula within the linear response theory:
\begin{subequations}\label{ohc}
	\begin{gather}
		\sigma_{\textrm{OH(SH)}}  = \frac{e}{\hbar} \sum_{n} \int \frac{ d^{3} \mathbf{k} }{ (2 \pi)^{3} } f_{n\mathbf{k}}  \Omega_{n}^{X_{z}}(\mathbf{k}),  \label{eq:kubo} \\
		\Omega_{n}^{X_{z}}(\mathbf{k}) = 2\hbar^{2} \sum_{m \neq n} \mathrm{Im} \left[ \frac{ \bra{u_{n\mathbf{k}}} j_{y}^{X_{z}} \ket{u_{m\mathbf{k}}} \bra{u_{m\mathbf{k}}} v_{x} \ket{u_{n\mathbf{k}}} }{ ( E_{n\mathbf{k}} - E_{m\mathbf{k}} + i \eta )^{2} } \right], \label{eq:berry}
	\end{gather}
\end{subequations}
where $f_{n\mathbf{k}}$ is the Fermi-Dirac distribution function, $\ket{u_{n\mathbf{k}}} $ is a periodic part of the Bloch state whose energy eigenvalue is $E_{n \mathbf{k}}$, $ v_{x} $ is $x$-component of the velocity operator, and $ j_{y}^{X_{z}} $ is $y$-component of the orbital(spin) current operator $j_{y}^{X_{z}} = ( X_{z} v_{y} + v_{y} X_{z} )/2$ with $z$-component of the orbital(spin) angular momentum $X_z = L_z(S_z)$.

\section{Mechanism}\label{sec:mechanism}

The orbital-texture-based mechanism of the OHE and SHE was recently reported but it was illustrated mainly for $p$ orbital systems. In this section, we demonstrate that the orbital-texture-based mechanism presented in Ref.~\onlinecite{go} works for $d$ orbital systems as well. For demonstration, we consider Pt, one of the well-studied materials for its large SHC. \cite{kimura2007, guo2008, sagasta2016}

\subsection{OHE from the orbital texture}\label{subsec:mechanism1}

In a multi-orbital system, hybridization between different orbitals leads to variation of the orbital characters with respect to $\mathbf{k}$. For example, in Figs.~\ref{fig:orbital_character}(b)-(d), $d_{xy}$, $d_{yz}$, and $d_{zx}$ orbital characters of fcc Pt are shown on top of the Fermi surface. It can be clearly seen that the orbital characters change with $\mathbf{k}$. This $\mathbf{k}$-space orbital texture is ubiquitous in the most multi-orbital systems, even in centrosymmetric materials. From this observation, we can make a toy model. Assuming a spherical shape Fermi surface, one can model this situation as radial and tangential type orbital textures. For example, in the $k_x k_y$-plane, a state with the tangential orbital character is written as $\ket{u_{\mathbf{k}}^{\textup{(t)}}} = \cos\theta_\mathbf{k} \ket{d_{yz}} - \sin\theta_\mathbf{k} \ket{d_{zx}}$, where $\tan \theta_\mathbf{k} = k_y/k_x$. In Fig.~\ref{fig:ohe}(b), tangential orbitals are shown on top of the Fermi surface in the $k_x k_y$ plane. Note that the expectation value of the orbital angular momentum is zero for the tangential state in {\it equilibrium}, which is a manifestation of the orbital quenching.\cite{kittel2004} Application of an external electric field $\mathbf{E}=E_x \hat{\mathbf{x}}$ shifts the Fermi sea by $\Delta k_x = -eE_x\delta t/\hbar$. Since this process occurs in a time $\delta t$ shorter than the momentum relaxation time caused by impurity scattering, it is intrinsic. Since the shifted state in {\it nonequilibrium} is not an eigenstate, the orbital character starts to change, hybridizing with other orbital character bands, i.e. a radial character band, $\ket{u_{\mathbf{k}}^{\textup{(r)}}} = \sin\theta_\mathbf{k} \ket{d_{yz}} + \cos\theta_\mathbf{k} \ket{d_{zx}}$. As a result of the hybridization, the nonequilibrium state for each $\mathbf{k}$ acquires positive or negative $\langle L_{z} \rangle$ depending on whether its $k_{y}$ is positive or negative as shown in Fig.~\ref{fig:ohe}(b). Although total angular momentum is zero when summed over all occupied states due to spatial inversion and time-reversal symmetries, the orbital Hall current $\sim \langle L_{z} v_{y} +  v_{y} L_{z} \rangle/2 $ is finite. This illustrates the mechanism of the OHE based on the momentum-space orbital texture in $d$ orbital systems.

\subsection{OHC and SHC of Pt}\label{subsec:mechanism2}

\begin{figure}[t]
	\includegraphics[angle=0, width=0.47\textwidth]{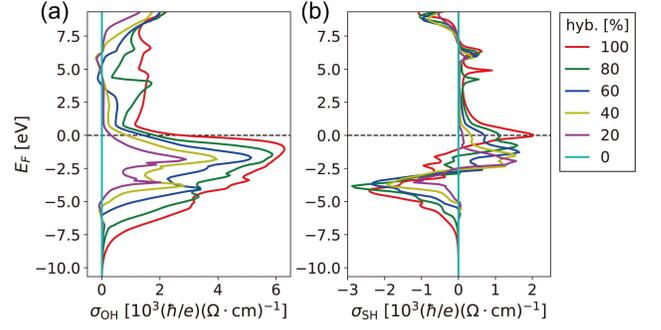}
	\caption{(a) OHC ($\sigma_{\textrm{OH}}$) and (b) SHC ($\sigma_{\textrm{SH}}$) in Pt as a function of the Fermi energy ($E_\mathrm{F}$) for different orbital hybridization strengths with respect to its full value (100\%).}
	\label{fig:pt1}
\end{figure}

As a prototypical example of the $d$ orbital system, we consider the intrinsic OHE and SHE of fcc Pt in detail. In Figs.~\ref{fig:pt1}(a) and \ref{fig:pt1}(b), the OHC and SHC are shown as a function of the Fermi energy $ E_{\mathrm{F}} $, respectively, for different orbital hybridization strengths with respect to its real value (100\%). Here, the orbital hybridization strength is controlled by varying orbital-mixing hoppings (see Appendix for details). We emphasize that the orbital hybridization strength is responsible for forming the orbital texture. If there is no orbital hybridization (0\%), we find that both OHC and SHC are zero because the orbital texture is absent [Fig.~\ref{fig:pt1}]. 

Red curves in Figs.~\ref{fig:pt1}(a) and \ref{fig:pt1}(b) show the OHC and SHC for 100\% orbital hybridization. In this case, we find that $\sigma_{\textrm{OH}} \approx 2700 \ ( \hbar/ e ) ( \Omega \cdot \mathrm{cm} )^{-1} $ and $\sigma_{\textrm{SH}} \approx 2000 \ ( \hbar/ e ) ( \Omega \cdot \mathrm{cm} )^{-1} $ for the true Fermi energy (horizontal dashed line). The large SHC in Pt is consistent with other theoretical calculations\cite{guo2008, tanaka2008} and experimental results,~\cite{sagasta2016, stamm2017} which verifies reliability of our calculation based on the TB model. 
Under gradual decrease of the orbital hybridization strength, we observe that both OHC and SHC monotonically decrease [Fig.~\ref{fig:pt1}]. From this, we conclude that the orbital-texture-based mechanism gives the most dominant contribution to the intrinsic OHE and SHE in Pt.\cite{note}
%
%

%

\begin{figure}[t]
	\includegraphics[angle=0, width=0.47\textwidth]{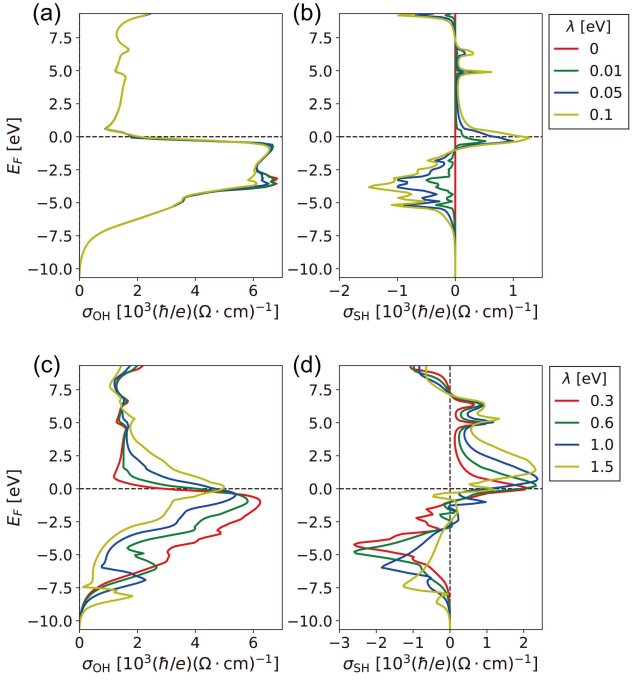}
	\caption{(a) OHC ($\sigma_\mathrm{OH}$) and (b) SHC ($\sigma_\mathrm{SH}$) of Pt for various SOC strengths $\lambda$ in small SOC regime. (c) $\sigma_\mathrm{OH}$ and (d) $\sigma_\mathrm{SH}$ in large SOC regime. The true value of SOC constant for Pt is $\lambda \approx 0.27\ \mathrm{eV}$.}
	\label{fig:pt2}
\end{figure}

The role of SOC is to convert the OHE into SHE. Thus, we investigate SOC dependence of the OHC and SHC, and their correlations. We vary $\lambda_d$ in Eq.~\eqref{eq:soc2} in our calculation because $\lambda_s$ does not contribute to SOC and $\lambda_p$ does not affect the result because $p$ states are at least $5\ \mathrm{eV}$ above the Fermi energy. In the below, we shortly write $\lambda_d$ as $\lambda$. Figures \ref{fig:pt2}(a) and \ref{fig:pt2}(b) respectively show the OHC and SHC when $\lambda$ is smaller than the real value $\lambda \approx 0.27$ eV. We find that in the small $\lambda$ regime the OHC barely changes, while the SHC increases monotonically with increasing SOC. Note that the OHE occurs in the absence of SOC, whereas the SHE does not. 
Meanwhile, Figs.~\ref{fig:pt2}(c) and \ref{fig:pt2}(d) respectively show the OHC and SHC when $\lambda$ is larger than the real value. In the large $\lambda$ regime, interestingly, the OHC experiences a drastic change. As $\lambda$ increases, the height of the peak for the OHC becomes smaller and the position of the peak shifts [Fig.~\ref{fig:pt2}(c)]. This is because the band structure is largely deformed by strong SOC. Consequently, as $\lambda$ increases, the OHC increases when $E_{\mathrm{F}} > 0$ and decreases when $E_{\mathrm{F}} < 0$. At true $E_{\mathrm{F}}$, which is zero, the OHC happens to be stable with respect to SOC. The Fermi energy being at this borderline appears to be accidental. For the SHC in the large SOC regime, the heights of the peaks of the SHC do not increase further even for larger $\lambda$ [Fig.~\ref{fig:pt2}(d)]. This is because a correlation between the orbital and spin saturates, forming a total angular momentum ($\mathbf{J} = \mathbf{L}+\mathbf{S}$) state. Although the heights of the peaks saturate, the SHC can show different behaviors at the Fermi energy because the positions of the peaks change.

\begin{figure}[t]
	\includegraphics[angle=0, width=0.40\textwidth]{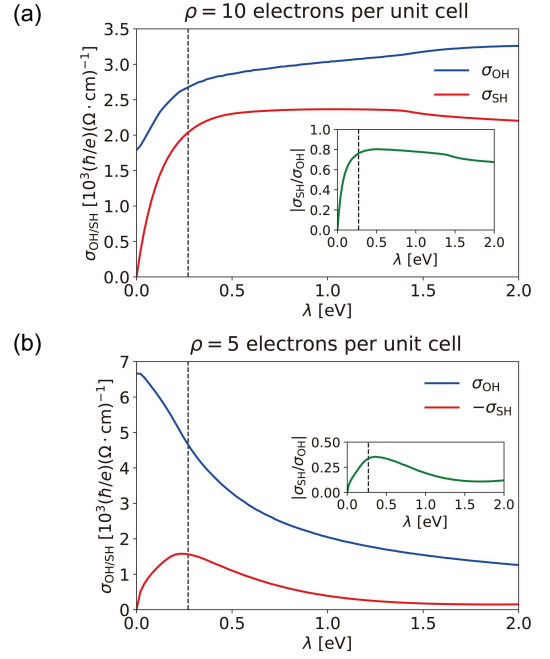}
	\caption{(a) OHC ($\sigma_\mathrm{OH}$) (blue line) and SHC ($\sigma_\mathrm{SH}$) (red line) of Pt as a function of the SOC constant $\lambda$ for a fixed valence electron density $\rho=10$ electrons per unit cell. The inset shows the ratio $|\sigma_\mathrm{SH}/\sigma_\mathrm{OH}|$. (b) Same plot for a fixed valence electron density $\rho=5$ electrons per unit cell. The vertical dashed line at $\lambda = 0.27$ eV indicates the true value of $\lambda$ in Pt.
	}
	\label{fig:pt3}
\end{figure}

Above results imply that large SOC does not necessarily guarantee larger SHC, and it depends on details of the band structure and the location of the Fermi energy. To illustrate this point, in Fig.~\ref{fig:pt3}(a), we plot the OHC and SHC with respect to $\lambda$ when the valence electron density is fixed to $\rho=10$ electrons per unit cell, which corresponds to true density of Pt. When $\lambda=0$, the SHC (red line) is zero but the OHC (blue line) is finite, which is expected because the OHE can occur even in the absence of SOC while the SHE requires SOC. 
As $\lambda$ increases from $0$, the SHC increases rapidly until $\lambda \approx 0.3\ \mathrm{eV}$, where it saturates to $\sim 2000\ (\hbar/e)(\Omega\cdot\mathrm{cm})^{-1}$. For Pt, true SOC constant (shown in black dashed line) $\lambda \approx 0.27\ \mathrm{eV}$ is located at the start of the saturation point. This explains large SHE of Pt observed from experiments.\cite{kimura2007, morota2011, sagasta2016} For exceptionally large SOC regime where $\lambda \gtrsim 1.0\ \mathrm{eV}$, however, the SHC starts to decrease weakly. On the other hand, the OHC keeps increasing as $\lambda$ increases although the increase is slow for $\lambda \gtrsim 0.3\ \mathrm{eV}$. In the inset, the ratio of the SHC over the OHC is shown, which can be interpreted as how much of the OHE is converted into the SHE by SOC. The conversion efficiency increases until $\lambda \approx 0.3\ \mathrm{eV}$ and saturates to $\sim 0.8$. For extremely large SOC regime, the conversion efficiency decreases slightly. However, SOC dependences of the OHC and SHC can differ depending on the Fermi energy. In Fig.~\ref{fig:pt3}(b), the OHC and SHC are shown for valence electron density $\rho=5$ electrons per unit cell. In this case, the OHC decreases dramatically for increasing $\lambda$. While the SHC increases monotonically in small SOC regime, it starts to diminish for large SOC because the SHE is converted from the OHE which is decreasing with $\lambda$.

\begin{figure}[t]
	\includegraphics[angle=0, width=0.45\textwidth]{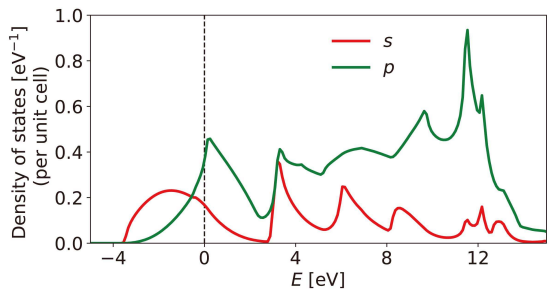}
	\caption{Orbital-resolved density of states for bcc Li. The red(green) line displays the density of states of $s$($p$) electrons. The vertical dashed line indicates the Fermi energy.} 
	\label{fig:dos_li}
\end{figure}

\section{OHE and SHE in weak SOC metals}\label{sec:light}

In the previous section, we showed that the OHE can arise in the presence of the momentum-space orbital texture, even without any aid from SOC. We also showed that the SHC increases rapidly for increasing SOC strength $\lambda$ especially when $\lambda$ is small. This implies that even when SOC is not significantly large, the OHC may be as large as the SHC of heavy metals, and the SHC may be non-negligible. In this section, we present results for metals with small SOC in body-centered cubic (bcc) (Li, V, Cr, Mn) and fcc (Al, Ni, Cu) structures. We present the results for $sp$ metals (Li, Al) and $sd$ metals (V, Cr, Mn, Ni, Cu) in order.

\begin{figure}[t]
	\includegraphics[angle=0, width=0.47\textwidth]{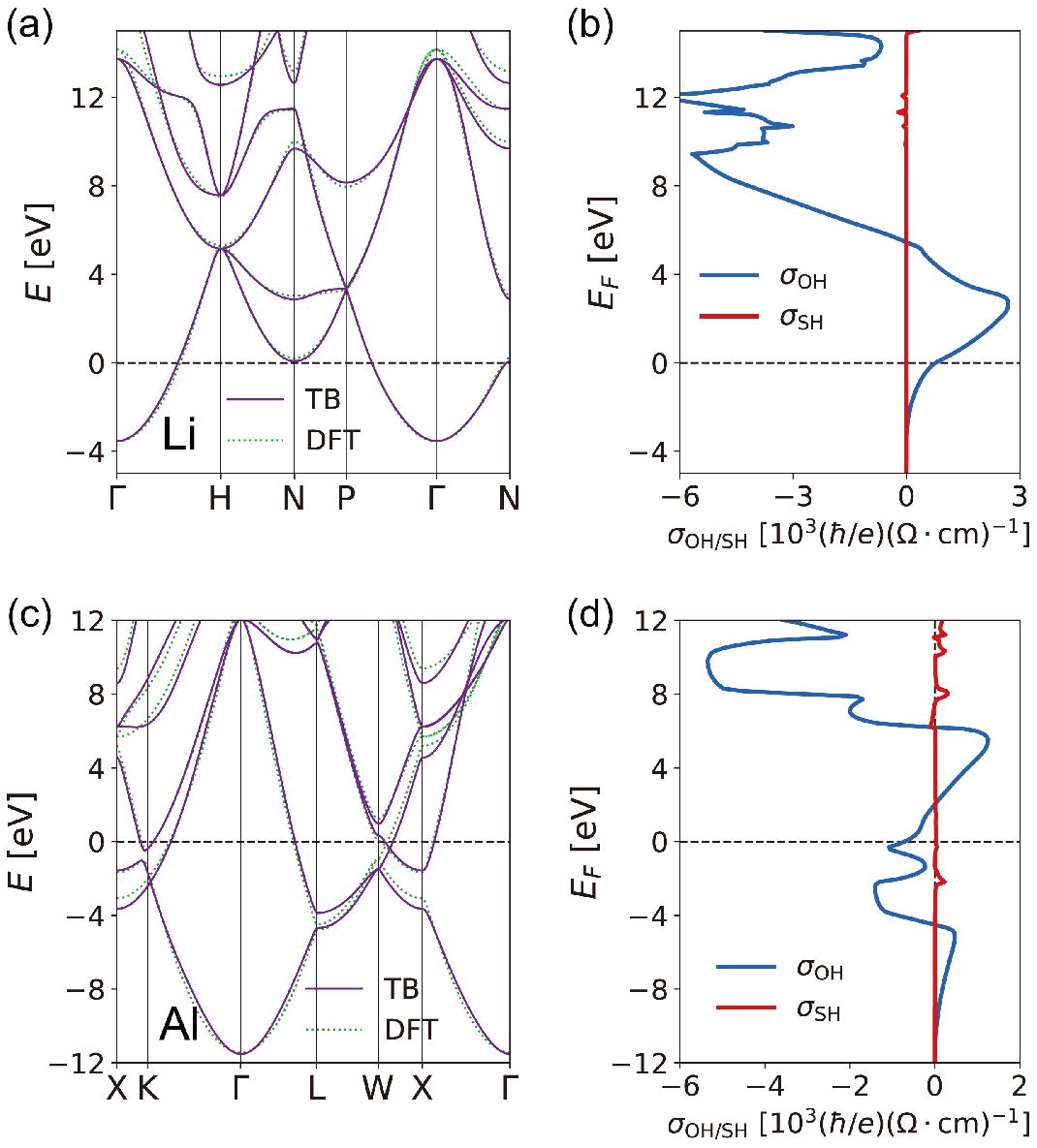}
	\caption{(a) Band structure for bcc Li calculated from the TB model (purple solid line) and DFT (green dotted line). (b) The Fermi energy ($E_\mathrm{F}$) dependence of the OHC ($\sigma_\mathrm{OH}$) (blue line) and SHC ($\sigma_\mathrm{OH}$) (red line) for bcc Li. The results for fcc Al are shown in (c) and (d).} 
	\label{fig:sp}
\end{figure}

\subsection{${sp}$ metals: Li, Al}\label{subsec:light1}

Although Li is an alkali metal, the $p$ character is as large as the $s$ character near the Fermi energy [Fig.~\ref{fig:dos_li}]. Thus, the $sp$ hybridization induces an orbital texture and the OHE can emerge.~\cite{go} Figures \ref{fig:sp}(a) and \ref{fig:sp}(b) show the band structure and the OHC and SHC for bcc Li. The SHC is almost zero due to very tiny size of SOC, while the OHC is large $\sigma_\mathrm{OH}\approx 800\ ( \hbar/ e ) ( \Omega \cdot \mathrm{cm} )^{-1} $. We attribute sizable magnitude of the OHC to the $sp$ hybridization.\cite{go} We notice that a peak of the OHC is located around $2\ \mathrm{eV}$ above the Fermi energy. This implies that the OHC can be further enhanced by finding a material having similar band structure with larger electron density, i.e. by alloying. Although Be has one more electron than Li, it cannot be such a material since it has a different crystal structure (hexagonal close-packed (hcp) structure) unfortunately and has qualitatively different band structure.

For Al, the band structure is shown in Fig.~\ref{fig:sp}(c), where there are $p$ character bands hybridizing with the $s$ character band (band bottom around $E=-11\ \mathrm{eV}$ near $\Gamma$) around the Fermi energy. The OHC and SHC for Al are displayed in Fig.~\ref{fig:sp}(d). Similar to the case of Li, the SHC is almost negligible $\sigma_\mathrm{SH}\approx 30\ ( \hbar/ e ) ( \Omega \cdot \mathrm{cm} )^{-1} $, while the OHC is large $\sigma_\mathrm{OH}\approx -700 \ ( \hbar/ e ) ( \Omega \cdot \mathrm{cm} )^{-1} $. Similar to Li, the large OHC of Al is attributed to the $sp$ hybridization. \cite{go}

\subsection{$sd$ metals: V, Cr, Mn, Ni, Cu}\label{subsec:light2}

\begin{figure}[t!]
	\includegraphics[angle=0, width=0.47\textwidth]{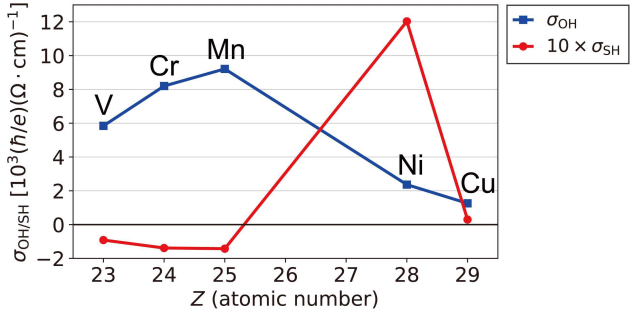}
	\caption{ OHC ($\sigma_\mathrm{OH}$) (blue square) and SHC ($\sigma_\mathrm{SH}$) (red circle) for several 3$d$ transition metals. $\sigma_\mathrm{SH}$ is enlarged by 10 times for the visibility of the data. Here V, Cr, Mn have bcc structures, and Ni (ferromagnetic) and Cu have fcc structures.} 
	\label{fig:3d}
\end{figure}

Most studies on the SHE so far have focused on $4d$ and $5d$ transition metals because large SOC in these materials tends to generate large SHE.\cite{tanaka2008, kontani2009, guo2008, freimuth2010, morota2011, sagasta2016} For $3d$ transition metals, on the other hand, theoretical study is limited~\cite{freimuth2010} despite the fact that several experiments have reported that 3$d$ transition metals such as V,\cite{wang2017} Cr,\cite{qu2015} Ni,\cite{du2014} and Py\cite{miao2013} exhibit large SHAs. This motivates us to study the OHE and SHE in $3d$ transition metals. We present results of the OHC and SHC for $sd$ metals, V, Cr, Mn in bcc structures and Ni, Cu in fcc structures. In these materials, the orbital texture arising from the $d$ orbital degree of freedom is crucial for the OHE as explained in Sec.~\ref{sec:mechanism}. In Fig.~\ref{fig:3d}, the OHC (blue square) and SHC (red circle) are shown for these elements in the increasing order of the atomic number $Z$. We find that the OHCs are in general gigantic $\sigma_\mathrm{OH} \sim 10^{3} - 10^{4} \ ( \hbar/ e ) ( \Omega \cdot \mathrm{cm} )^{-1}$. This result is similar to $4d$ and $5d$ transition metals.\cite{tanaka2008} Surprisingly, in spite of small SOC, $3d$ transition metals exhibit sizable magnitude of SHC, $\sigma_\mathrm{SH}\sim 10^2 ( \hbar/ e ) ( \Omega \cdot \mathrm{cm} )^{-1}$. Especially, the SHC of Ni is exceptionally large, $\sigma_\mathrm{SH}\sim 10^3 ( \hbar/ e ) ( \Omega \cdot \mathrm{cm} )^{-1}$, which is comparable to the SHC of Pt. Its origin will be explained in detail in the Discussion section.

\begin{figure}[t]	
	\includegraphics[angle=0, width=0.47\textwidth]{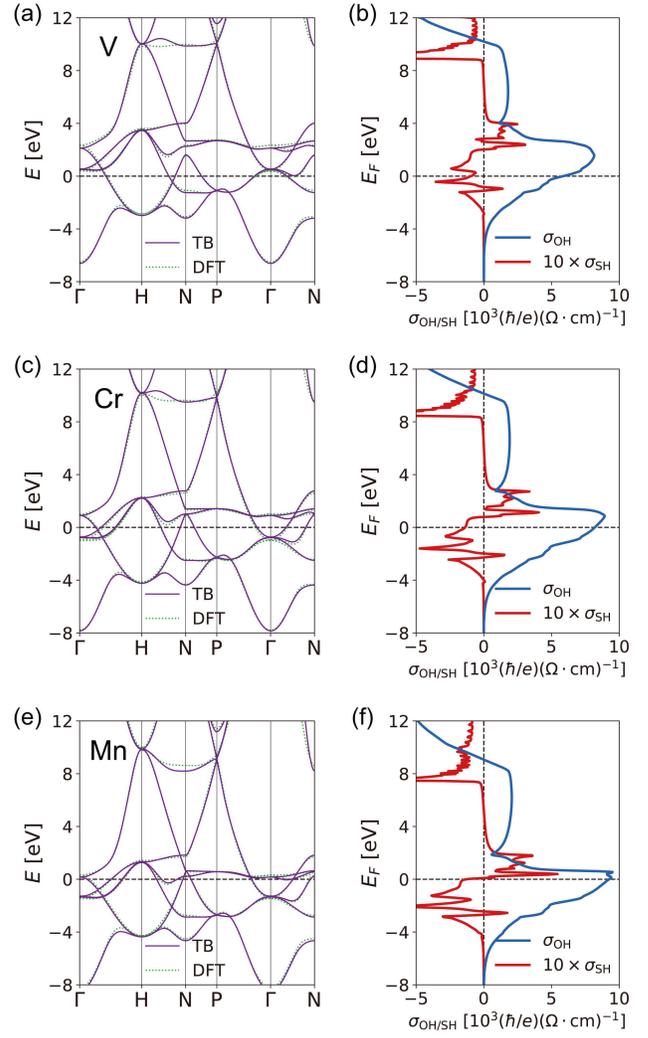}
	\caption{(a) Band structure for bcc V calculated by TB model (purple solid line) and DFT (green dotted line). (b) The Fermi energy ($E_\mathrm{F}$) dependence of the OHC ($\sigma_\mathrm{OH}$) (blue line) and SHC ($\sigma_\mathrm{SH}$) (red line) for V. The results for bcc Cr are represented in (c) and (d), and for bcc Mn in (e) and (f). Note that $\sigma_\mathrm{SH}$ is enlarged 10 times for the visibility. } 
	\label{fig:sd1}
\end{figure}

\begin{figure}[t]
	\includegraphics[angle=0, width=0.47\textwidth]{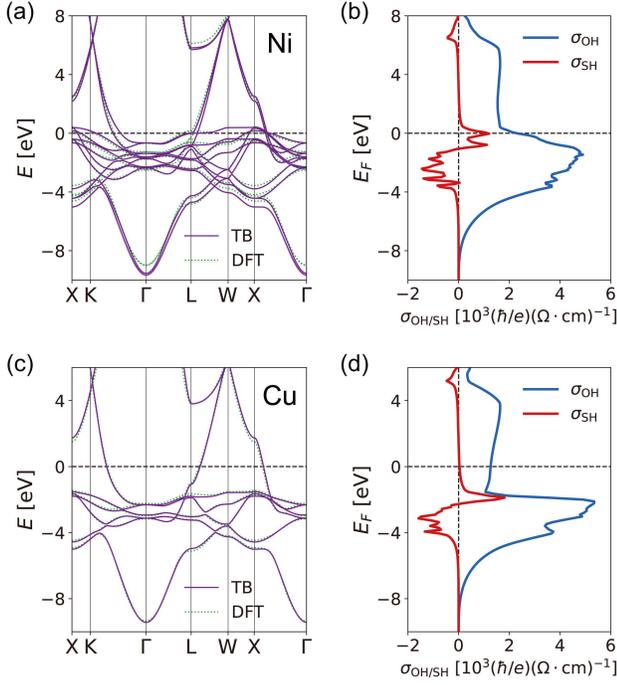}
	\caption{(a) Band structure for fcc Ni (ferromagnetic) calculated by TB model (purple solid line) and DFT (green dotted line). (b) The Fermi energy ($E_\mathrm{F}$) dependence of the OHC ($\sigma_\mathrm{OH}$) (blue line) and SHC ($\sigma_\mathrm{SH}$) (red line) for Ni. The results for fcc Cu are shown in (c) and (d).} 
	\label{fig:sd2}
\end{figure}

First, we present the results for V, Cr, and Mn, which are adjacent to each other in the periodic table. We assume bcc structure for these elements. Although Mn can have $\alpha$-Mn structure, bcc structure can also be stabilized. While Cr shows antiferromagnetic ordering at room temperature with the N\'{e}el temperature $T_\mathrm{N}=311$ K,\cite{ashcroft} we assume a paramagnetic phase because a recent experiment has reported that the SHE in Cr is independent of the antiferromagnetic ordering.\cite{qu2015} The band structures of V, Cr, and Mn, and their OHCs and SHCs as a function of the Fermi energy are shown in Fig.~\ref{fig:sd1}. It is interesting to see that overall shapes of the band structures [Figs.~\ref{fig:sd1}(a), \ref{fig:sd1}(c), \ref{fig:sd1}(e)] and the Fermi energy dependences of the OHC and SHC [Figs.~\ref{fig:sd1}(b), \ref{fig:sd1}(d), \ref{fig:sd1}(f)] resemble each other. This is because valence electron orbitals and their hoppings are almost similar. The major difference between V, Cr, and Mn is the electron density and the location of the Fermi energy. We find that OHC for V is surprisingly large $\sigma_\mathrm{OH}\approx 5800 \ ( \hbar/ e ) ( \Omega \cdot \mathrm{cm} )^{-1}$. However, its Fermi energy is still below the peak. Thus, it can be further enhanced when the Fermi energy increases about $2\ \mathrm{eV}$. Since Cr and Mn have more valence electrons than V, the OHC at the Fermi energy is approaching the peak [Figs.~\ref{fig:sd1}(d) and \ref{fig:sd1}(f)]. The OHCs in these elements are gigantic, $\sigma_{\textrm{OH}} \approx 8200$ and $9200\ ( \hbar/ e ) ( \Omega \cdot \mathrm{cm} )^{-1} $ for Cr and Mn, respectively. These values are 4$\sim$5 times larger than the SHC of Pt. Although the SHCs in these elements are much smaller than the OHCs, magnitude of the OHC is not negligible thus experimentally measurable. The values of SHCs are $\sigma_\mathrm{SH}\approx -90$, $-130$, $-130\ ( \hbar/ e ) ( \Omega \cdot \mathrm{cm} )^{-1}$ for V, Cr, Mn, respectively.

For fcc crystals, we consider Ni and Cu. For Ni, we assume a ferromagnetic phase since its Curie temperature is $T_\mathrm{C} = 627$ K, \cite{ashcroft} which is above the room temperature. In Fig.~\ref{fig:sd2}, the band structures and Fermi energy dependences of the OHC ans SHC are shown for Ni and Cu. Since both elements have similar valence orbitals, their band structures are quite close to each other except for the exchange splitting in the band structure of Ni [Figs.~\ref{fig:sd2}(a) and \ref{fig:sd2}(c)]. The OHCs are large for these elements, $\sigma_{\textrm{OH}}\approx$ 2300 and 1200 $ ( \hbar/ e ) ( \Omega \cdot \mathrm{cm} )^{-1} $ for Ni and Cu, respectively. Interestingly, Ni exhibits unexpectedly large SHC $\sigma_{\textrm{SH}} \approx$ 1200 $ ( \hbar/ e ) ( \Omega \cdot \mathrm{cm} )^{-1} $, which is about the same order of magnitude as Pt. On the other hand, Cu has much smaller SHC $\sigma_{\textrm{SH}} \approx$ 30 $ ( \hbar/ e ) ( \Omega \cdot \mathrm{cm} )^{-1} $. These results can be understood in the same manner as the case of bcc crystals. The OHC and SHC of Ni are larger than those of Cu, because the Fermi energy of Ni is closer to the peak positions of the OHC and SHC.

\begin{figure}[t]
	\includegraphics[angle=0, width=0.45\textwidth]{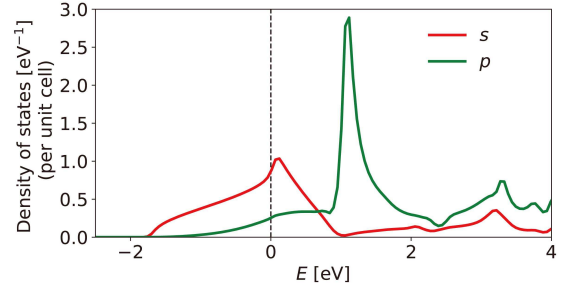}
	\caption{Orbital-resolved density of states per unit cell for Cs. The red(green) line displays the density of states of $s$($p$) electrons. The vertical dashed line indicates the Fermi energy.} 
	\label{fig:dos_cs}
\end{figure}

\section{Discussion}\label{sec:discussion}

In this paper, the orbital angular momentum was defined around the atomic center. This choice of coordinate origin is natural when there is a single atom in a unit cell, but for multi-atom cases, other choices are possible. Although the orbital angular momentum depends on the choice of origin, the OHC is independent of that choice in centrosymmetric systmes. To see this, suppose we shift the coordinate $\mathbf{r} \mapsto \mathbf{r} + \delta \mathbf{r}$, then the orbital angular momentum would be changed by $\delta \mathbf{L} = \delta \mathbf{r} \times \mathbf{p}$. This gives an extra contribution to the OHC [Eq. \eqref{eq:kubo}],
\begin{equation}\label{origin}
	\delta \sigma_{\textrm{OH}} = \frac{e}{\hbar} \sum_{n} \int \frac{ d^{3} \mathbf{k} }{ (2 \pi)^{3} } f_{n\mathbf{k}}  \delta \Omega_{n}^{L_{z}}(\mathbf{k}),
\end{equation}
where $\delta \Omega_{n}^{L_{z}}(\mathbf{k})$ includes terms proportional to $(\delta r_x \bra{u_{n\mathbf{k}}} v_{y}^{2} \ket{u_{m\mathbf{k}}} - \delta r_y \bra{u_{n\mathbf{k}}} v_{x}v_{y} \ket{u_{m\mathbf{k}}}  )\bra{u_{m\mathbf{k}}} v_{x} \ket{u_{n\mathbf{k}}}$. Note that the velocity operator is odd under the inversion operation and $\delta \Omega_{n}^{L_{z}}(\mathbf{k})$ is odd function of $\mathbf{k}$. Therefore, the integration over $\mathbf{k}$ in Eq.~\eqref{origin} is zero in centrosymmetric materials, and the OHC is independent of the origin choice. Thus, with more than one atoms in a unit cell, one may simply adds all the contribution to the orbital angular momentum operator from each atom and employ the same formula in Eq.~\eqref{ohc}.

Now let us comment on the evaluation of the orbital angular momentum operator. In the TB calculation presented in this paper, we employed the matrix representation in Eqs.~\eqref{orbital} and \eqref{orbital2} within the tight-binding model using the cubic harmonic basis. In DFT calculation, it can be calculated as 
\begin{equation}\label{oam}
\mathbf{L} = \sum_{\mathbf{R}} \sum_{n,m} 
\ket{\phi_{n\mathbf{R}}} \bra{\phi_{n\mathbf{R}}} 
(\mathbf{r} - \mathbf{R}) \times \mathbf{p}
\ket{\phi_{m\mathbf{R}}} \bra{\phi_{m\mathbf{R}}},
\end{equation}
by using localized Wannier function $\ket{\phi_{n\mathbf{R}}}$ where $n$ and $m$ are orbital indices and $\mathbf{R}$ is the Bravais lattice vector.\cite{go2017} In full-potential linearized augmented plane wave scheme (FLAPW),\cite{wimmer1981} where the muffin-tin is defined, one may simply evaluate $\mathbf{r} \times \mathbf{p}$ and integrate it over a region inside the muffin-tin.\cite{hanke2016}

We have shown that the intrinsic OHE exists in many multi-orbital systems even though SOC is small. However, the orbital texture cannot be formed in the $s$ orbital system, thus there is no OHE in materials where the $s$ orbital character dominates near the Fermi energy. As an illustration, we consider Cs. In Fig.~\ref{fig:dos_cs}, orbital-resolved density of states for Cs is shown. We see that major orbital character at the Fermi energy is coming from the $s$ orbital. For this reason, the OHC for Cs is much smaller,  $\sigma_{\textrm{OH}} \approx 180 \ ( \hbar/ e ) ( \Omega \cdot \mathrm{cm} )^{-1} $, compared to the materials we considered in this paper. Consequently, the SHC is nearly zero as well. 

\begin{table}[t]
	\begin{tabular}
		{cccc}\hline \hline
		Element  & $\theta_\textrm{OH}$ & $\theta_\textrm{SH}$ & $\theta_\textrm{SH}^{\textup{exp}}$ (Ref.~\onlinecite{du2014}) \\ \hline
		bcc V 
		& 
		$3.4$ 
		& 
		$-5.3 \times 10^{-2}$ 
		&
		$-(1.0 \pm 0.1) \times 10^{-2}$
		\\
		bcc Cr 
		& 
		$14 $ 
		&
		$-2.3 \times 10^{-1}$
		&
		$-(5.1 \pm 0.5)\times 10^{-2}$ 
		\\
		bcc Mn 
		& 
		$18 $ 
		& 
		$-2.5 \times 10^{-1}$ 
		&
		$-(1.9 \pm 0.1) \times 10^{-3} $
		\\
		fcc Cu 
		&
		$1.6 \times 10^{-2}$ 
		&
		\ \ \ $3.8 \times 10^{-4}$ 
		&
		\ \ \ $(3.2 \pm 0.3)\times 10^{-3}$
		\\
		\hline \hline
	\end{tabular}
	\caption{OHA ($\theta_\mathrm{OH}$) and SHA ($\theta_\mathrm{SH}$) for V, Cr, Mn, Cu calculated from the TB model. Experimentally measured values of the resistivity and SHA $\theta_\mathrm{SH}^{\textup{exp}}$ are taken from Ref.~\onlinecite{du2014} 
	}
	\label{table}
\end{table}

In Sec.~\ref{subsec:light2}, the OHC and SHC of $3d$ transition metals were presented [Fig.~\ref{fig:3d}].  Here, we compare our results with experimental data in Ref.~\onlinecite{du2014} where the SHAs of $3d$ transition metals were measured by spin pumping experiment. The Hund's rule type behavior of SHA in Fig.~3(b) in Ref. \onlinecite{du2014} is similar to our calculated values of the SHC in Fig.~\ref{fig:3d}. For example, V, Cr, Mn have negative SHCs, while Ni and Cr have positive SHCs, which is consistent with experimental observation.
For quantative comparison, we compute the orbital Hall angle (OHA) and SHA as $ \theta_\textrm{OH}= \sigma_\textrm{OH} \rho (2e/\hbar)$ and $ \theta_\textrm{SH}= \sigma_\textrm{SH} \rho (2e/\hbar) $, respectively. The values for the resistivity $\rho$ are taken from  Ref.~\onlinecite{du2014}. In Table \ref{table}, calculated values for OHAs (second column) and SHAs (third column) are shown together with experimentally measured values of SHAs from Ref.~\onlinecite{du2014} (fourth column).  The SHAs calculated from the theory are larger than experimental values except for Cu.\cite{du2014} Both theory and experiment agree that the SHA of Cr is about 5 times larger than SHC of V. In case of Mn, the comparison with the experiment is tricky since our calculation in Figs.~\ref{fig:sd1}(e) and \ref{fig:sd1}(f) is for bcc structure whereas the experiment was probably performed for the $\alpha$-Mn structure. If we ignore this diffrence, there is a discrepancy between calculation and experiment.

\begin{figure}[t]
	\includegraphics[angle=0, width=0.47\textwidth]{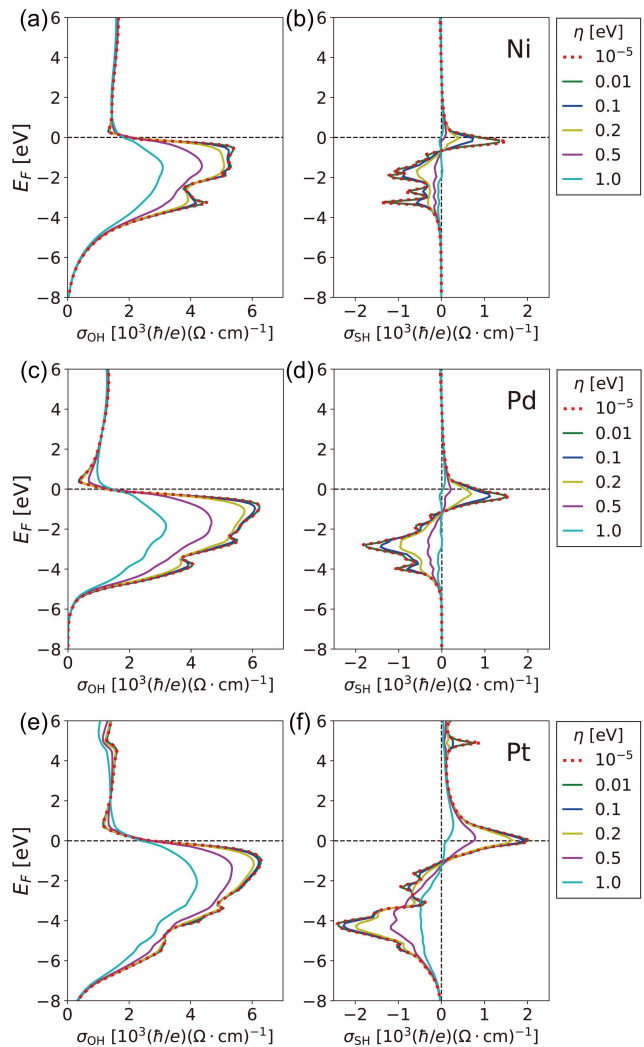}
	\caption{(a) OHC ($\sigma_\mathrm{OH}$) and (b) SHC ($\sigma_\mathrm{SH}$) of paramagnetic fcc Ni for various $\eta$ in Eq. \eqref{eq:kubo}. The results for fcc Pd are represented in (c) and (d), and for fcc Pt in (e) and (f). These three elements are in the same column in the periodic table.}
	\label{fig:ni_pd_pt}
\end{figure}

References \onlinecite{miao2013, du2014} reported exceptionally large SHAs of Ni and Py (which contains 80\% of Ni and 20\% of Fe), about half of the SHA of Pt. In our calculation, we have also seen that the SHC of Ni is large, $\sigma_\mathrm{SH}\approx 1200\ (\hbar/e)(\Omega\cdot\mathrm{cm})^{-1}$, which is about half of the SHC of Pt. In order to understand the origin of the large SHE in Ni, we compare Ni with Pd and Pt, which are in the same column of the periodic table and are all in fcc structure. The red dotted lines in Figs.~\ref{fig:ni_pd_pt}(a)-(f) show the OHCs and SHCs of Ni, Pd, and Pt. For the comparison, we calculate the OHCs and SHCs of Ni, Pd, Pt in a paramagnetic phase. Since the band structures are very similar to each other, their OHCs are very close to each other [Figs.~\ref{fig:ni_pd_pt}(a), \ref{fig:ni_pd_pt}(c), and \ref{fig:ni_pd_pt}(e)]. On the other hand, SOC strengths are different,  $\lambda = 0.045 \ \textrm{eV}$, $0.092 \ \textrm{eV}$, $0.27 \ \textrm{eV}$, for Ni, Pd, Pt, respectively. Thus, the SHC of Ni is smaller than the SHC of Pt. Nonetheless, the heights of the peak at the Fermi energy do not differ much [Figs.~\ref{fig:ni_pd_pt}(b), \ref{fig:ni_pd_pt}(d), and \ref{fig:ni_pd_pt}(f)]. Especially, SHCs of Ni and Pd are almost similar. Thus, we conclude that with increasing SOC strength the SHC starts to saturate from the SOC strength of Ni. In fact, Ni and Pd can be considered as the weak SOC versions of Pt. The results of Pt for $\lambda = 0.05 \ \textrm{eV}$ and $0.1 \ \textrm{eV}$ [Figs.~\ref{fig:pt2}(a) and \ref{fig:pt2}(b)] are very similar to those of Ni and Pd, respectively. Moreover, electron density is optimal, where the Fermi energy is located at the peak of the SHC for Ni, Pd, Pt. These make the intrinsic SHE of Ni, Pd, Pt large.

However, for Ni, the width of the SHC peak near the Fermi energy is narrower than the corresponding SHC of Pt due to smaller energy splitting caused by weak SOC. Then the SHC would be more unstable against peak broadening from impurity scatterings. In order to check this, we calculate the OHCs and SHCs of Ni, Pd, and Pt by replacing $\eta$ in Eq.~\eqref{eq:kubo} to finite values. It corresponds to a self-energy correction of the OHC or SHC in the presence of disorder potential in the scale of $\eta$. Figures \ref{fig:ni_pd_pt}(a)-(f) show that the OHC and SHC peaks are reduced in general as $\eta$ increases. However, the OHC of Ni is stable due to the accidental position of its Fermi energy, as shown in Fig.~\ref{fig:ni_pd_pt}(a). Additionally, Pd and Pt show the similar behaviors of the OHCs with respect to $\eta$, regardless of their difference in the SOC strength [Figs.~\ref{fig:ni_pd_pt}(c) and \ref{fig:ni_pd_pt}(e)]. On the other hand, the SHC peak is more largely reduced by $\eta$ especially when SOC strength is smaller because the width of the SHC peak is narrower. For example, the heights of the SHC peaks near the Fermi energy are lowered by about half when $\eta$ goes down to $0.1\ \textrm{eV}$, $0.2\ \textrm{eV}$, and $0.5\ \textrm{eV}$ for Ni, Pd, and Pt, respectively [Figs.~\ref{fig:ni_pd_pt}(b), \ref{fig:ni_pd_pt}(d), and \ref{fig:ni_pd_pt}(f)]. This tendency is in line with their SOC strengths, $\lambda = 0.045 \ \textrm{eV}$, $0.092 \ \textrm{eV}$, $0.27 \ \textrm{eV}$. Therefore, the SHCs of Ni, Pd, and Pt decrease with increasing $\eta$, and Ni shows the largest decrease among them. Another source of $\eta$ is the electron-phonon scattering. At room temperature, $\eta$ due to the electron-phonon scattering is comparable to the thermal energy $k_{\mathrm{B}}T = 0.025 \ \mathrm{eV}$. Considering that the peak width of the SHC of Ni is at least 10 times broader than $k_\mathrm{B}T$, the large SHC of Ni can be stable with respect to the electron-phonon scattering. We found that the SHC is still large, $\sigma_\mathrm{SH}\approx 990\ (\hbar/e)(\Omega\cdot\mathrm{cm})^{-1}$ for $\eta = 0.025\ \mathrm{eV}$, which is about 20\% smaller than the SHC obtained for $\eta \mapsto 0^{+}$. Therefore, the large SHC of Ni can be measured in room temperature experiments, provided the crystalline ordering in Ni is sufficiently good.

\begin{figure}[t]
	\includegraphics[angle=0, width=0.47\textwidth]{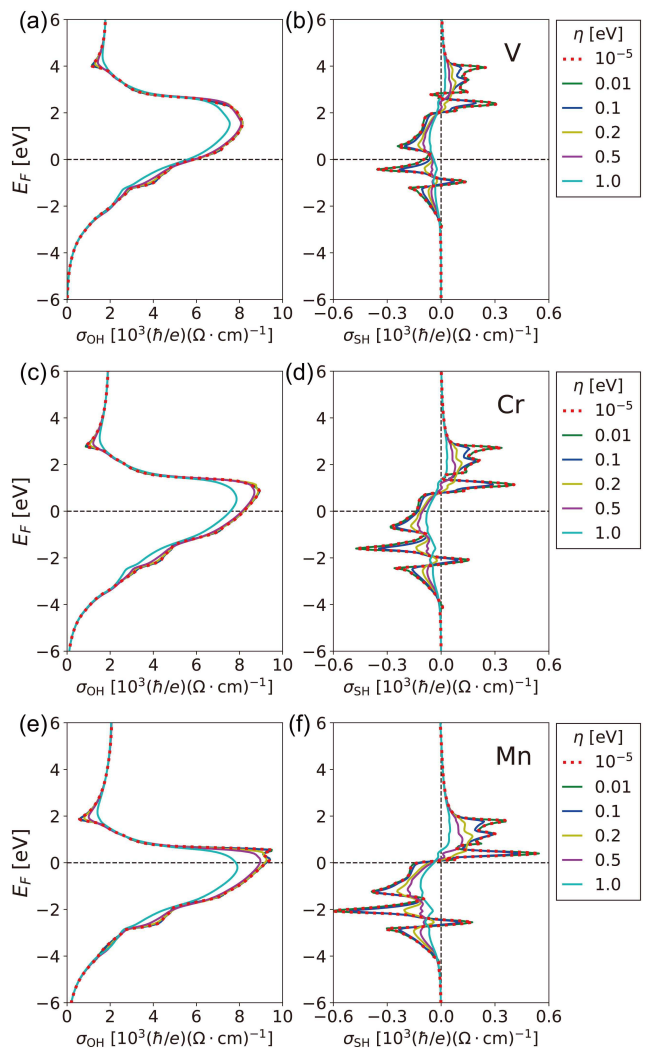}
	\caption{(a) OHC ($\sigma_\mathrm{OH}$) and (b) SHC ($\sigma_\mathrm{SH}$) of bcc V for various $\eta$ in Eq. \eqref{eq:kubo}. The results for bcc Cr are represented in (c) and (d), and for bcc Mn in (e) and (f). These three elements are in the same row in the periodic table.}
	\label{fig:v_cr_mn}
\end{figure}

Similarly, we calculate the OHCs and SHCs of bcc V, Cr, and Mn for different $\eta$'s [Figs.~\ref{fig:v_cr_mn}(a)-(f)]. The OHCs and SHCs of these elements also decrease with increasing $\eta$, but the OHCs maintain the large values even when $\eta=1.0 \ \textrm{eV}$ [Figs.~\ref{fig:v_cr_mn}(a), \ref{fig:v_cr_mn}(c), and \ref{fig:v_cr_mn}(e)]. Figures \ref{fig:v_cr_mn}(b), \ref{fig:v_cr_mn}(d), and \ref{fig:v_cr_mn}(f) show that the narrow peaks of the SHC curves are largely lowered by $\eta$. However, in contrast with the former cases, the Fermi energy is not positioned near the peaks of the SHC, thus the true SHC is not directly determined by the peak heights. For instance, the SHC peaks almost vanish when $\eta=0.5 \ \textrm{eV}$, but the SHCs near the Fermi energy for V and Cr are relatively robust [Figs.~\ref{fig:v_cr_mn}(b) and \ref{fig:v_cr_mn}(d)]. For Mn, on the other hand, the SHC near the true Fermi energy experiences the rapid change with respect to the Fermi energy, thus it is rather unstable against $\eta$ [Fig.~\ref{fig:v_cr_mn}(f)].

Recent experiments demonstrated that the SHE in Cu can be significantly enhanced by surface oxidation.\cite{an2016, gao2018} This is surprising because the SHC of unoxidized Cu is small $\sigma_\mathrm{SH}\approx 30\ (\hbar/e)(\Omega\cdot\mathrm{cm})^{-1}$ according to our calculation. One possible factor that may contribute to the experimental results\cite{an2016, gao2018} is the Fermi level lowering by the oxidation. Upon such lowering, the electronic configuration of Cu becomes similar to Ni, where the peak of the SHC is located at the Fermi energy. For instance, if Cu atom loses 0.5 electron per unit cell, then the Fermi level would lie near $-1.5\ \mathrm{eV}$ below the true Fermi energy [Fig.~\ref{fig:sd2}(c)]. Then, the SHC becomes $\sigma_{\textrm{SH}} \approx 150\ ( \hbar/ e ) ( \Omega \cdot \mathrm{cm} )^{-1} $ [Fig.~\ref{fig:sd2}(d)]. Since SOC strengths of Cu and Ni are similar to each other, the peak heights of the SHC are also similar. Therefore, we expect that novel engineering of band filling by oxidization would provide a route to enhancing the OHE and SHE.

Finally, we discuss how to observe the OHE in experiments. The most direct way of measuring the OHE(SHE) would be to probe the boundary accumulation of the orbital(spin) moments. For separating the signals from the spin and orbital, x-ray circular magnetic dichroism measurement and the sum rule can be applied. \cite{obrien1994, bonetti2017} The magneto-optical Kerr effect (MOKE) can also be used for measuring the OHE similar to the measurement of the SHE.\cite{kato2004, stamm2017, puebla2017, liu2018} However, MOKE cannot distinguish the signals from the orbital and spin. In this sense, experiments on materials having small SOC have great advantage. In these materials, the OHE dominates over the SHE and it is unlikely that the SHE gives any significant contribution, thus one could detect the OHE more easily compared to strong SOC materials where both OHE and SHE coexist.

\section{Summary}\label{sec:summary}

We have demonstrated that the orbital-texture-based mechanism for the intrinsic OHE applies to real materials as well. For illustration, we considered Pt and found that the orbital hybridization is crucial for generating the OHE and SHE since it is responsible for forming the orbital texture. By varying SOC strength in Pt, we found that while the SHC is proportional to SOC strength in the small SOC regime, larger SOC does not necessarily enhance the SHC in the large SOC regime. The reason is that large SOC may decrease the OHE by modifying the band structure thereby reducing the orbital texture. As well as Pt, we considered many metallic systems having small SOC, Li and Al as examples of the $sp$ metal and V, Cr, Mn, Ni, Cu as examples of the $sd$ metal. We found gigantic OHCs in these materials. For $sd$ metals, we found that the SHCs are not negligible although they are smaller than OHCs. This could explain the large SHAs observed from experiments on many $3d$ transition metals.\cite{wang2017, qu2015, du2014, miao2013, an2016, gao2018}
Our finding not only enriches understanding of the intrinsic OHE and SHE, but also widens material choice to weak SOC materials for spintronics. Moreover, our microscopic analysis based on the electronic  band structure provides a route to enhancing these effects, i.e. by tuning the electron density with oxidization. Since the orbital degree of freedom is ubiquitous in solids, utilizing the orbital degree of freedom in transport phenomena will open a plethora of \emph{orbitronics}.\cite{bernevig2005b}

\begin{acknowledgments}
D.J. was supported by Global Ph.D. Fellowship Program by National Research Foundation of Korea (Grant No. 2018H1A2A1060270). D.G. was supported by Global Ph.D. Fellowship Program by National Research Foundation of Korea (Grant No. 2014H1A2A1019219 ). D.J., D.G., and H.-W.L. were supported by the SSTF (Grant No. BA-1501-07).
\end{acknowledgments}

\appendix*
\section{Details of the calculation}

We employed the TB model based on linear combinations of atomic orbitals (LCAO)\cite{slater1954} to obtain the energy bands and corresponding wave functions. An electron wave function can be expanded in terms of Bloch-like states $ \ket{\varphi_{n \sigma \mathbf{k}}} $ given by 
\begin{equation}\label{eq:bloch}
	\ket{\varphi_{n \sigma \mathbf{k}}} = \frac{1}{\sqrt{N}} \sum_{\mathbf{R}} e^{i\mathbf{k} \cdot \mathbf{R}} \ket{\phi_{n \sigma \mathbf{R}}},
\end{equation}
where $n = s, p_{x}, p_{y}, p_{z}, d_{z^2}, d_{x^{2}-y^{2}}, d_{xy}, d_{yz}, d_{zx}$ denotes the orbital, $\sigma=\uparrow, \downarrow$ is the spin, and $\mathbf{R}$ is the Bravais lattice whose number is $N$. With these orthogonal basis states, we constructed $18 \times 18$ TB Hamiltonian matrix whose element $\bra{\varphi_{n' \sigma' \mathbf{k}}} H \ket{\varphi_{n \sigma \mathbf{k}}} $ is expressed in terms of on-site energies and hopping amplitudes. Within the two-center approximation, these parameters were taken from Ref. \onlinecite{handbook}. We took into account up to the second nearest hoppings for fcc structure, and the third nearest hoppings for bcc structure. For the integration in Eq. \eqref{eq:kubo}, we used 150 $\times$ 150 $\times$ 150 $\mathbf{k}$-mesh grid, where we found less than 1\% change of the OHC and SHC compared to the integration on 200 $\times$ 200 $\times$ 200 $\mathbf{k}$-mesh grid.

In Sec.~\ref{subsec:mechanism2}, we manipulated orbital hybridization strength of Pt by controlling the hopping parameters in TB model. Here, we did not vary the on-site energies and the hoppings between the same orbitals such as $6s$-$6s$ hopping, $6p_{i}$-$6p_{i}$ hopping ($i=x, y, z$), and $5d_{i}$-$5d_{i}$ hopping ($i=z^{2}, x^{2}-y^{2}, xy, yz, zx$). We only controlled the hopping parameters between different orbitals, which are responsible for the orbital texture. For example, the green curves in Fig.~\ref{fig:pt1} indicate the OHC and SHC when the hopping parameters between different orbitals are reduced to 80\% of their real values. This process was carried out by scaling the off-diagonal elements of Hamiltonian. The reduction of orbital hybridization can be implemented in other ways, e.g. scaling the Slater-Koster parameters. Depending on the ways, the quantitative values of the OHC and SHC with reduced orbital hybridization can be different. Nevertheless, when the orbital hybridization is zero, all the off-diagonal terms of Hamiltonian vanish equivalently, so the trends in Fig.~\ref{fig:pt1} should be unchanged.

The SOC constant $\lambda_{l}$ in Eq. \eqref{eq:soc2} was determined by fitting the energy bands obtained from the TB model to those obtained from DFT. For example, in 3$d$ transition metals, three-fold degenerated $t_{2g}$ levels at $\Gamma$ point split into two $J_{\textrm{eff}}=3/2$ levels and one $J_{\textrm{eff}}=1/2$ level by SOC.\cite{kim2008} We determined the value of SOC constant $\lambda_{d}$ so that the energy splitting at $\Gamma$ point
\begin{align}\label{eq:splitting}
	\Delta_{\textrm{SOC}} = \bra{J_{\textrm{eff}} = 3/2} \frac{2\lambda_{d}}{\hbar^{2}} ( \mathbf{L} \cdot \mathbf{S} ) \ket{J_{\textrm{eff}} = 3/2} \nonumber \\
	- \bra{J_{\textrm{eff}} = 1/2} \frac{2\lambda_{d}}{\hbar^{2}} ( \mathbf{L} \cdot \mathbf{S} ) \ket{J_{\textrm{eff}} = 1/2}
\end{align}
fits into the DFT result.

We performed self-consistent DFT calculations of the electronic structure for each element using \texttt{FLEUR} code, \cite{fleur} which implements the FLAPW method.\cite{wimmer1981} Exchange and correlation effects were treated within the generalized gradient approximation.\cite{perdew1996} We sampled the irreducible Brillouin zone using $16\times 16 \times 16$ $\mathbf{k}$-mesh grid. SOC was included self-consistently within the second-variation scheme. \cite{li1990}


\begin{thebibliography}{50}%
	\makeatletter
	\providecommand \@ifxundefined [1]{%
		\@ifx{#1\undefined}
	}%
	\providecommand \@ifnum [1]{%
		\ifnum #1\expandafter \@firstoftwo
		\else \expandafter \@secondoftwo
		\fi
	}%
	\providecommand \@ifx [1]{%
		\ifx #1\expandafter \@firstoftwo
		\else \expandafter \@secondoftwo
		\fi
	}%
	\providecommand \natexlab [1]{#1}%
	\providecommand \enquote  [1]{``#1''}%
	\providecommand \bibnamefont  [1]{#1}%
	\providecommand \bibfnamefont [1]{#1}%
	\providecommand \citenamefont [1]{#1}%
	\providecommand \href@noop [0]{\@secondoftwo}%
	\providecommand \href [0]{\begingroup \@sanitize@url \@href}%
	\providecommand \@href[1]{\@@startlink{#1}\@@href}%
	\providecommand \@@href[1]{\endgroup#1\@@endlink}%
	\providecommand \@sanitize@url [0]{\catcode `\\12\catcode `\$12\catcode
		`\&12\catcode `\#12\catcode `\^12\catcode `\_12\catcode `\%12\relax}%
	\providecommand \@@startlink[1]{}%
	\providecommand \@@endlink[0]{}%
	\providecommand \url  [0]{\begingroup\@sanitize@url \@url }%
	\providecommand \@url [1]{\endgroup\@href {#1}{\urlprefix }}%
	\providecommand \urlprefix  [0]{URL }%
	\providecommand \Eprint [0]{\href }%
	\providecommand \doibase [0]{http://dx.doi.org/}%
	\providecommand \selectlanguage [0]{\@gobble}%
	\providecommand \bibinfo  [0]{\@secondoftwo}%
	\providecommand \bibfield  [0]{\@secondoftwo}%
	\providecommand \translation [1]{[#1]}%
	\providecommand \BibitemOpen [0]{}%
	\providecommand \bibitemStop [0]{}%
	\providecommand \bibitemNoStop [0]{.\EOS\space}%
	\providecommand \EOS [0]{\spacefactor3000\relax}%
	\providecommand \BibitemShut  [1]{\csname bibitem#1\endcsname}%
	\let\auto@bib@innerbib\@empty
	\bibitem [{\citenamefont {{D'Yakonov}}\ and\ \citenamefont
		{{Perel'}}(1971)}]{dyakonov1971}%
	\BibitemOpen
	\bibfield  {author} {\bibinfo {author} {\bibfnamefont {M.~I.}\ \bibnamefont
			{{D'Yakonov}}}\ and\ \bibinfo {author} {\bibfnamefont {V.~I.}\ \bibnamefont
			{{Perel'}}},\ }\href@noop {} {\bibfield  {journal} {\bibinfo  {journal} {JETP
				Lett.}\ }\textbf {\bibinfo {volume} {13}},\ \bibinfo {pages} {467} (\bibinfo
		{year} {1971})}\BibitemShut {NoStop}%
	\bibitem [{\citenamefont {Hirsch}(1999)}]{hirsch1999}%
	\BibitemOpen
	\bibfield  {author} {\bibinfo {author} {\bibfnamefont {J.~E.}\ \bibnamefont
			{Hirsch}},\ }\href {\doibase 10.1103/PhysRevLett.83.1834} {\bibfield
		{journal} {\bibinfo  {journal} {Phys. Rev. Lett.}\ }\textbf {\bibinfo
			{volume} {83}},\ \bibinfo {pages} {1834} (\bibinfo {year}
		{1999})}\BibitemShut {NoStop}%
	\bibitem [{\citenamefont {Engel}\ \emph {et~al.}(2006)\citenamefont {Engel},
		\citenamefont {Rashba},\ and\ \citenamefont {Halperin}}]{engel2007}%
	\BibitemOpen
	\bibfield  {author} {\bibinfo {author} {\bibfnamefont {H.-A.}\ \bibnamefont
			{Engel}}, \bibinfo {author} {\bibfnamefont {E.~I.}\ \bibnamefont {Rashba}}, \
		and\ \bibinfo {author} {\bibfnamefont {B.~I.}\ \bibnamefont {Halperin}},\
	}in\ \href@noop {} {\emph {\bibinfo {booktitle} {Handbook of Magnetism and
				Advanced Magnetic Materials}}},\ Vol.~\bibinfo {volume} {5},\ \bibinfo
	{editor} {edited by\ \bibinfo {editor} {\bibfnamefont {H.}~\bibnamefont
			{Kronm\"uller}}\ and\ \bibinfo {editor} {\bibfnamefont {S.}~\bibnamefont
			{Parkin}}}\ (\bibinfo  {publisher} {Wiley},\ \bibinfo {address}
	{Chichester},\ \bibinfo {year} {2006})\ p.\ \bibinfo {pages}
	{2858}\BibitemShut {NoStop}%
	\bibitem [{\citenamefont {Schliemann}(2006)}]{schliemann2006}%
	\BibitemOpen
	\bibfield  {author} {\bibinfo {author} {\bibfnamefont {J.}~\bibnamefont
			{Schliemann}},\ }\href {\doibase 10.1142/S021797920603370X} {\bibfield
		{journal} {\bibinfo  {journal} {Int. J. Mod. Phys. B}\ }\textbf {\bibinfo
			{volume} {20}},\ \bibinfo {pages} {1015} (\bibinfo {year}
		{2006})}\BibitemShut {NoStop}%
	\bibitem [{\citenamefont {Sinova}\ \emph {et~al.}(2015)\citenamefont {Sinova},
		\citenamefont {Valenzuela}, \citenamefont {Wunderlich}, \citenamefont
		{Back},\ and\ \citenamefont {Jungwirth}}]{sinova2015}%
	\BibitemOpen
	\bibfield  {author} {\bibinfo {author} {\bibfnamefont {J.}~\bibnamefont
			{Sinova}}, \bibinfo {author} {\bibfnamefont {S.~O.}\ \bibnamefont
			{Valenzuela}}, \bibinfo {author} {\bibfnamefont {J.}~\bibnamefont
			{Wunderlich}}, \bibinfo {author} {\bibfnamefont {C.~H.}\ \bibnamefont
			{Back}}, \ and\ \bibinfo {author} {\bibfnamefont {T.}~\bibnamefont
			{Jungwirth}},\ }\href {\doibase 10.1103/RevModPhys.87.1213} {\bibfield
		{journal} {\bibinfo  {journal} {Rev. Mod. Phys.}\ }\textbf {\bibinfo {volume}
			{87}},\ \bibinfo {pages} {1213} (\bibinfo {year} {2015})}\BibitemShut
	{NoStop}%
	\bibitem [{\citenamefont {Saitoh}\ \emph {et~al.}(2006)\citenamefont {Saitoh},
		\citenamefont {Ueda}, \citenamefont {Miyajima},\ and\ \citenamefont
		{Tatara}}]{saitoh2006}%
	\BibitemOpen
	\bibfield  {author} {\bibinfo {author} {\bibfnamefont {E.}~\bibnamefont
			{Saitoh}}, \bibinfo {author} {\bibfnamefont {M.}~\bibnamefont {Ueda}},
		\bibinfo {author} {\bibfnamefont {H.}~\bibnamefont {Miyajima}}, \ and\
		\bibinfo {author} {\bibfnamefont {G.}~\bibnamefont {Tatara}},\ }\href
	{\doibase 10.1063/1.2199473} {\bibfield  {journal} {\bibinfo  {journal}
			{Appl. Phys. Lett.}\ }\textbf {\bibinfo {volume} {88}},\ \bibinfo {pages}
		{182509} (\bibinfo {year} {2006})}\BibitemShut {NoStop}%
	\bibitem [{\citenamefont {Kimura}\ \emph {et~al.}(2007)\citenamefont {Kimura},
		\citenamefont {Otani}, \citenamefont {Sato}, \citenamefont {Takahashi},\ and\
		\citenamefont {Maekawa}}]{kimura2007}%
	\BibitemOpen
	\bibfield  {author} {\bibinfo {author} {\bibfnamefont {T.}~\bibnamefont
			{Kimura}}, \bibinfo {author} {\bibfnamefont {Y.}~\bibnamefont {Otani}},
		\bibinfo {author} {\bibfnamefont {T.}~\bibnamefont {Sato}}, \bibinfo {author}
		{\bibfnamefont {S.}~\bibnamefont {Takahashi}}, \ and\ \bibinfo {author}
		{\bibfnamefont {S.}~\bibnamefont {Maekawa}},\ }\href {\doibase
		10.1103/PhysRevLett.98.156601} {\bibfield  {journal} {\bibinfo  {journal}
			{Phys. Rev. Lett.}\ }\textbf {\bibinfo {volume} {98}},\ \bibinfo {pages}
		{156601} (\bibinfo {year} {2007})}\BibitemShut {NoStop}%
	\bibitem [{\citenamefont {Niimi}\ and\ \citenamefont
		{Otani}(2015)}]{niimi2015}%
	\BibitemOpen
	\bibfield  {author} {\bibinfo {author} {\bibfnamefont {Y.}~\bibnamefont
			{Niimi}}\ and\ \bibinfo {author} {\bibfnamefont {Y.}~\bibnamefont {Otani}},\
	}\href {http://stacks.iop.org/0034-4885/78/i=12/a=124501} {\bibfield
		{journal} {\bibinfo  {journal} {Rep. Prog. Phys.}\ }\textbf {\bibinfo
			{volume} {78}},\ \bibinfo {pages} {124501} (\bibinfo {year}
		{2015})}\BibitemShut {NoStop}%
	\bibitem [{\citenamefont {Liu}\ \emph {et~al.}(2012{\natexlab{a}})\citenamefont
		{Liu}, \citenamefont {Lee}, \citenamefont {Gudmundsen}, \citenamefont
		{Ralph},\ and\ \citenamefont {Buhrman}}]{liu2012prl}%
	\BibitemOpen
	\bibfield  {author} {\bibinfo {author} {\bibfnamefont {L.}~\bibnamefont
			{Liu}}, \bibinfo {author} {\bibfnamefont {O.~J.}\ \bibnamefont {Lee}},
		\bibinfo {author} {\bibfnamefont {T.~J.}\ \bibnamefont {Gudmundsen}},
		\bibinfo {author} {\bibfnamefont {D.~C.}\ \bibnamefont {Ralph}}, \ and\
		\bibinfo {author} {\bibfnamefont {R.~A.}\ \bibnamefont {Buhrman}},\ }\href
	{\doibase 10.1103/PhysRevLett.109.096602} {\bibfield  {journal} {\bibinfo
			{journal} {Phys. Rev. Lett.}\ }\textbf {\bibinfo {volume} {109}},\ \bibinfo
		{pages} {096602} (\bibinfo {year} {2012}{\natexlab{a}})}\BibitemShut
	{NoStop}%
	\bibitem [{\citenamefont {Liu}\ \emph {et~al.}(2012{\natexlab{b}})\citenamefont
		{Liu}, \citenamefont {Pai}, \citenamefont {Li}, \citenamefont {Tseng},
		\citenamefont {Ralph},\ and\ \citenamefont {Buhrman}}]{liu2012science}%
	\BibitemOpen
	\bibfield  {author} {\bibinfo {author} {\bibfnamefont {L.}~\bibnamefont
			{Liu}}, \bibinfo {author} {\bibfnamefont {C.-F.}\ \bibnamefont {Pai}},
		\bibinfo {author} {\bibfnamefont {Y.}~\bibnamefont {Li}}, \bibinfo {author}
		{\bibfnamefont {H.~W.}\ \bibnamefont {Tseng}}, \bibinfo {author}
		{\bibfnamefont {D.~C.}\ \bibnamefont {Ralph}}, \ and\ \bibinfo {author}
		{\bibfnamefont {R.~A.}\ \bibnamefont {Buhrman}},\ }\href {\doibase
		10.1126/science.1218197} {\bibfield  {journal} {\bibinfo  {journal}
			{Science}\ }\textbf {\bibinfo {volume} {336}},\ \bibinfo {pages} {555}
		(\bibinfo {year} {2012}{\natexlab{b}})}\BibitemShut {NoStop}%
	\bibitem [{\citenamefont {Tanaka}\ \emph {et~al.}(2008)\citenamefont {Tanaka},
		\citenamefont {Kontani}, \citenamefont {Naito}, \citenamefont {Naito},
		\citenamefont {Hirashima}, \citenamefont {Yamada},\ and\ \citenamefont
		{Inoue}}]{tanaka2008}%
	\BibitemOpen
	\bibfield  {author} {\bibinfo {author} {\bibfnamefont {T.}~\bibnamefont
			{Tanaka}}, \bibinfo {author} {\bibfnamefont {H.}~\bibnamefont {Kontani}},
		\bibinfo {author} {\bibfnamefont {M.}~\bibnamefont {Naito}}, \bibinfo
		{author} {\bibfnamefont {T.}~\bibnamefont {Naito}}, \bibinfo {author}
		{\bibfnamefont {D.~S.}\ \bibnamefont {Hirashima}}, \bibinfo {author}
		{\bibfnamefont {K.}~\bibnamefont {Yamada}}, \ and\ \bibinfo {author}
		{\bibfnamefont {J.}~\bibnamefont {Inoue}},\ }\href {\doibase
		10.1103/PhysRevB.77.165117} {\bibfield  {journal} {\bibinfo  {journal} {Phys.
				Rev. B}\ }\textbf {\bibinfo {volume} {77}},\ \bibinfo {pages} {165117}
		(\bibinfo {year} {2008})}\BibitemShut {NoStop}%
	\bibitem [{\citenamefont {Kontani}\ \emph {et~al.}(2009)\citenamefont
		{Kontani}, \citenamefont {Tanaka}, \citenamefont {Hirashima}, \citenamefont
		{Yamada},\ and\ \citenamefont {Inoue}}]{kontani2009}%
	\BibitemOpen
	\bibfield  {author} {\bibinfo {author} {\bibfnamefont {H.}~\bibnamefont
			{Kontani}}, \bibinfo {author} {\bibfnamefont {T.}~\bibnamefont {Tanaka}},
		\bibinfo {author} {\bibfnamefont {D.~S.}\ \bibnamefont {Hirashima}}, \bibinfo
		{author} {\bibfnamefont {K.}~\bibnamefont {Yamada}}, \ and\ \bibinfo {author}
		{\bibfnamefont {J.}~\bibnamefont {Inoue}},\ }\href {\doibase
		10.1103/PhysRevLett.102.016601} {\bibfield  {journal} {\bibinfo  {journal}
			{Phys. Rev. Lett.}\ }\textbf {\bibinfo {volume} {102}},\ \bibinfo {pages}
		{016601} (\bibinfo {year} {2009})}\BibitemShut {NoStop}%
	\bibitem [{\citenamefont {Guo}\ \emph {et~al.}(2008)\citenamefont {Guo},
		\citenamefont {Murakami}, \citenamefont {Chen},\ and\ \citenamefont
		{Nagaosa}}]{guo2008}%
	\BibitemOpen
	\bibfield  {author} {\bibinfo {author} {\bibfnamefont {G.~Y.}\ \bibnamefont
			{Guo}}, \bibinfo {author} {\bibfnamefont {S.}~\bibnamefont {Murakami}},
		\bibinfo {author} {\bibfnamefont {T.-W.}\ \bibnamefont {Chen}}, \ and\
		\bibinfo {author} {\bibfnamefont {N.}~\bibnamefont {Nagaosa}},\ }\href
	{\doibase 10.1103/PhysRevLett.100.096401} {\bibfield  {journal} {\bibinfo
			{journal} {Phys. Rev. Lett.}\ }\textbf {\bibinfo {volume} {100}},\ \bibinfo
		{pages} {096401} (\bibinfo {year} {2008})}\BibitemShut {NoStop}%
	\bibitem [{\citenamefont {Freimuth}\ \emph {et~al.}(2010)\citenamefont
		{Freimuth}, \citenamefont {Bl\"ugel},\ and\ \citenamefont
		{Mokrousov}}]{freimuth2010}%
	\BibitemOpen
	\bibfield  {author} {\bibinfo {author} {\bibfnamefont {F.}~\bibnamefont
			{Freimuth}}, \bibinfo {author} {\bibfnamefont {S.}~\bibnamefont {Bl\"ugel}},
		\ and\ \bibinfo {author} {\bibfnamefont {Y.}~\bibnamefont {Mokrousov}},\
	}\href {\doibase 10.1103/PhysRevLett.105.246602} {\bibfield  {journal}
		{\bibinfo  {journal} {Phys. Rev. Lett.}\ }\textbf {\bibinfo {volume} {105}},\
		\bibinfo {pages} {246602} (\bibinfo {year} {2010})}\BibitemShut {NoStop}%
	\bibitem [{\citenamefont {Morota}\ \emph {et~al.}(2011)\citenamefont {Morota},
		\citenamefont {Niimi}, \citenamefont {Ohnishi}, \citenamefont {Wei},
		\citenamefont {Tanaka}, \citenamefont {Kontani}, \citenamefont {Kimura},\
		and\ \citenamefont {Otani}}]{morota2011}%
	\BibitemOpen
	\bibfield  {author} {\bibinfo {author} {\bibfnamefont {M.}~\bibnamefont
			{Morota}}, \bibinfo {author} {\bibfnamefont {Y.}~\bibnamefont {Niimi}},
		\bibinfo {author} {\bibfnamefont {K.}~\bibnamefont {Ohnishi}}, \bibinfo
		{author} {\bibfnamefont {D.~H.}\ \bibnamefont {Wei}}, \bibinfo {author}
		{\bibfnamefont {T.}~\bibnamefont {Tanaka}}, \bibinfo {author} {\bibfnamefont
			{H.}~\bibnamefont {Kontani}}, \bibinfo {author} {\bibfnamefont
			{T.}~\bibnamefont {Kimura}}, \ and\ \bibinfo {author} {\bibfnamefont
			{Y.}~\bibnamefont {Otani}},\ }\href {\doibase 10.1103/PhysRevB.83.174405}
	{\bibfield  {journal} {\bibinfo  {journal} {Phys. Rev. B}\ }\textbf {\bibinfo
			{volume} {83}},\ \bibinfo {pages} {174405} (\bibinfo {year}
		{2011})}\BibitemShut {NoStop}%
	\bibitem [{\citenamefont {Sagasta}\ \emph {et~al.}(2016)\citenamefont
		{Sagasta}, \citenamefont {Omori}, \citenamefont {Isasa}, \citenamefont
		{Gradhand}, \citenamefont {Hueso}, \citenamefont {Niimi}, \citenamefont
		{Otani},\ and\ \citenamefont {Casanova}}]{sagasta2016}%
	\BibitemOpen
	\bibfield  {author} {\bibinfo {author} {\bibfnamefont {E.}~\bibnamefont
			{Sagasta}}, \bibinfo {author} {\bibfnamefont {Y.}~\bibnamefont {Omori}},
		\bibinfo {author} {\bibfnamefont {M.}~\bibnamefont {Isasa}}, \bibinfo
		{author} {\bibfnamefont {M.}~\bibnamefont {Gradhand}}, \bibinfo {author}
		{\bibfnamefont {L.~E.}\ \bibnamefont {Hueso}}, \bibinfo {author}
		{\bibfnamefont {Y.}~\bibnamefont {Niimi}}, \bibinfo {author} {\bibfnamefont
			{Y.}~\bibnamefont {Otani}}, \ and\ \bibinfo {author} {\bibfnamefont
			{F.}~\bibnamefont {Casanova}},\ }\href {\doibase 10.1103/PhysRevB.94.060412}
	{\bibfield  {journal} {\bibinfo  {journal} {Phys. Rev. B}\ }\textbf {\bibinfo
			{volume} {94}},\ \bibinfo {pages} {060412} (\bibinfo {year}
		{2016})}\BibitemShut {NoStop}%
	\bibitem [{\citenamefont {Sinova}\ \emph {et~al.}(2004)\citenamefont {Sinova},
		\citenamefont {Culcer}, \citenamefont {Niu}, \citenamefont {Sinitsyn},
		\citenamefont {Jungwirth},\ and\ \citenamefont {MacDonald}}]{sinova2004}%
	\BibitemOpen
	\bibfield  {author} {\bibinfo {author} {\bibfnamefont {J.}~\bibnamefont
			{Sinova}}, \bibinfo {author} {\bibfnamefont {D.}~\bibnamefont {Culcer}},
		\bibinfo {author} {\bibfnamefont {Q.}~\bibnamefont {Niu}}, \bibinfo {author}
		{\bibfnamefont {N.~A.}\ \bibnamefont {Sinitsyn}}, \bibinfo {author}
		{\bibfnamefont {T.}~\bibnamefont {Jungwirth}}, \ and\ \bibinfo {author}
		{\bibfnamefont {A.~H.}\ \bibnamefont {MacDonald}},\ }\href {\doibase
		10.1103/PhysRevLett.92.126603} {\bibfield  {journal} {\bibinfo  {journal}
			{Phys. Rev. Lett.}\ }\textbf {\bibinfo {volume} {92}},\ \bibinfo {pages}
		{126603} (\bibinfo {year} {2004})}\BibitemShut {NoStop}%
	\bibitem [{\citenamefont {Murakami}\ \emph {et~al.}(2003)\citenamefont
		{Murakami}, \citenamefont {Nagaosa},\ and\ \citenamefont {Zhang}}]{murakami}%
	\BibitemOpen
	\bibfield  {author} {\bibinfo {author} {\bibfnamefont {S.}~\bibnamefont
			{Murakami}}, \bibinfo {author} {\bibfnamefont {N.}~\bibnamefont {Nagaosa}}, \
		and\ \bibinfo {author} {\bibfnamefont {S.-C.}\ \bibnamefont {Zhang}},\ }\href
	{\doibase 10.1126/science.1087128} {\bibfield  {journal} {\bibinfo  {journal}
			{Science}\ }\textbf {\bibinfo {volume} {301}},\ \bibinfo {pages} {1348}
		(\bibinfo {year} {2003})}\BibitemShut {NoStop}%
	\bibitem [{\citenamefont {Yao}\ and\ \citenamefont {Fang}(2005)}]{yao2005}%
	\BibitemOpen
	\bibfield  {author} {\bibinfo {author} {\bibfnamefont {Y.}~\bibnamefont
			{Yao}}\ and\ \bibinfo {author} {\bibfnamefont {Z.}~\bibnamefont {Fang}},\
	}\href {\doibase 10.1103/PhysRevLett.95.156601} {\bibfield  {journal}
		{\bibinfo  {journal} {Phys. Rev. Lett.}\ }\textbf {\bibinfo {volume} {95}},\
		\bibinfo {pages} {156601} (\bibinfo {year} {2005})}\BibitemShut {NoStop}%
	\bibitem [{\citenamefont {Kittel}(2004)}]{kittel2004}%
	\BibitemOpen
	\bibfield  {author} {\bibinfo {author} {\bibfnamefont {C.}~\bibnamefont
			{Kittel}},\ }\href {https://books.google.co.kr/books?id=kym4QgAACAAJ} {\emph
		{\bibinfo {title} {Introduction to Solid State Physics}}}\ (\bibinfo
	{publisher} {Wiley},\ \bibinfo {address} {New York},\ \bibinfo {year}
	{2004})\BibitemShut {NoStop}%
	\bibitem [{\citenamefont {Murakami}\ \emph {et~al.}(2004)\citenamefont
		{Murakami}, \citenamefont {Nagaosa},\ and\ \citenamefont
		{Zhang}}]{murakami2004prl}%
	\BibitemOpen
	\bibfield  {author} {\bibinfo {author} {\bibfnamefont {S.}~\bibnamefont
			{Murakami}}, \bibinfo {author} {\bibfnamefont {N.}~\bibnamefont {Nagaosa}}, \
		and\ \bibinfo {author} {\bibfnamefont {S.-C.}\ \bibnamefont {Zhang}},\ }\href
	{\doibase 10.1103/PhysRevLett.93.156804} {\bibfield  {journal} {\bibinfo
			{journal} {Phys. Rev. Lett.}\ }\textbf {\bibinfo {volume} {93}},\ \bibinfo
		{pages} {156804} (\bibinfo {year} {2004})}\BibitemShut {NoStop}%
	\bibitem [{\citenamefont {Go}\ \emph {et~al.}(2018)\citenamefont {Go},
		\citenamefont {Jo}, \citenamefont {Kim},\ and\ \citenamefont {Lee}}]{go}%
	\BibitemOpen
	\bibfield  {author} {\bibinfo {author} {\bibfnamefont {D.}~\bibnamefont
			{Go}}, \bibinfo {author} {\bibfnamefont {D.}~\bibnamefont {Jo}}, \bibinfo
		{author} {\bibfnamefont {C.}~\bibnamefont {Kim}}, \ and\ \bibinfo {author}
		{\bibfnamefont {H.-W.}\ \bibnamefont {Lee}},\ }\href {\doibase
		10.1103/PhysRevLett.121.086602} {\bibfield  {journal} {\bibinfo  {journal}
			{Phys. Rev. Lett.}\ }\textbf {\bibinfo {volume} {121}},\ \bibinfo {pages}
		{086602} (\bibinfo {year} {2018})}\BibitemShut {NoStop}%
	\bibitem [{\citenamefont {Kontani}\ \emph {et~al.}(2008)\citenamefont
		{Kontani}, \citenamefont {Tanaka}, \citenamefont {Hirashima}, \citenamefont
		{Yamada},\ and\ \citenamefont {Inoue}}]{kontani2008}%
	\BibitemOpen
	\bibfield  {author} {\bibinfo {author} {\bibfnamefont {H.}~\bibnamefont
			{Kontani}}, \bibinfo {author} {\bibfnamefont {T.}~\bibnamefont {Tanaka}},
		\bibinfo {author} {\bibfnamefont {D.~S.}\ \bibnamefont {Hirashima}}, \bibinfo
		{author} {\bibfnamefont {K.}~\bibnamefont {Yamada}}, \ and\ \bibinfo {author}
		{\bibfnamefont {J.}~\bibnamefont {Inoue}},\ }\href {\doibase
		10.1103/PhysRevLett.100.096601} {\bibfield  {journal} {\bibinfo  {journal}
			{Phys. Rev. Lett.}\ }\textbf {\bibinfo {volume} {100}},\ \bibinfo {pages}
		{096601} (\bibinfo {year} {2008})}\BibitemShut {NoStop}%
	\bibitem [{\citenamefont {Thonhauser}\ \emph {et~al.}(2005)\citenamefont
		{Thonhauser}, \citenamefont {Ceresoli}, \citenamefont {Vanderbilt},\ and\
		\citenamefont {Resta}}]{thonhauser2005}%
	\BibitemOpen
	\bibfield  {author} {\bibinfo {author} {\bibfnamefont {T.}~\bibnamefont
			{Thonhauser}}, \bibinfo {author} {\bibfnamefont {D.}~\bibnamefont
			{Ceresoli}}, \bibinfo {author} {\bibfnamefont {D.}~\bibnamefont
			{Vanderbilt}}, \ and\ \bibinfo {author} {\bibfnamefont {R.}~\bibnamefont
			{Resta}},\ }\href {\doibase 10.1103/PhysRevLett.95.137205} {\bibfield
		{journal} {\bibinfo  {journal} {Phys. Rev. Lett.}\ }\textbf {\bibinfo
			{volume} {95}},\ \bibinfo {pages} {137205} (\bibinfo {year}
		{2005})}\BibitemShut {NoStop}%
	\bibitem [{\citenamefont {Xiao}\ \emph {et~al.}(2005)\citenamefont {Xiao},
		\citenamefont {Shi},\ and\ \citenamefont {Niu}}]{xiao2005}%
	\BibitemOpen
	\bibfield  {author} {\bibinfo {author} {\bibfnamefont {D.}~\bibnamefont
			{Xiao}}, \bibinfo {author} {\bibfnamefont {J.}~\bibnamefont {Shi}}, \ and\
		\bibinfo {author} {\bibfnamefont {Q.}~\bibnamefont {Niu}},\ }\href {\doibase
		10.1103/PhysRevLett.95.137204} {\bibfield  {journal} {\bibinfo  {journal}
			{Phys. Rev. Lett.}\ }\textbf {\bibinfo {volume} {95}},\ \bibinfo {pages}
		{137204} (\bibinfo {year} {2005})}\BibitemShut {NoStop}%
	\bibitem [{\citenamefont {Wang}\ \emph {et~al.}(2017)\citenamefont {Wang},
		\citenamefont {Wang}, \citenamefont {Xie}, \citenamefont {Warsi},
		\citenamefont {Wu}, \citenamefont {Chen}, \citenamefont {Lorenz},
		\citenamefont {Fan},\ and\ \citenamefont {Xiao}}]{wang2017}%
	\BibitemOpen
	\bibfield  {author} {\bibinfo {author} {\bibfnamefont {T.}~\bibnamefont
			{Wang}}, \bibinfo {author} {\bibfnamefont {W.}~\bibnamefont {Wang}}, \bibinfo
		{author} {\bibfnamefont {Y.}~\bibnamefont {Xie}}, \bibinfo {author}
		{\bibfnamefont {M.~A.}\ \bibnamefont {Warsi}}, \bibinfo {author}
		{\bibfnamefont {J.}~\bibnamefont {Wu}}, \bibinfo {author} {\bibfnamefont
			{Y.}~\bibnamefont {Chen}}, \bibinfo {author} {\bibfnamefont {V.~O.}\
			\bibnamefont {Lorenz}}, \bibinfo {author} {\bibfnamefont {X.}~\bibnamefont
			{Fan}}, \ and\ \bibinfo {author} {\bibfnamefont {J.~Q.}\ \bibnamefont
			{Xiao}},\ }\href {\doibase 10.1038/s41598-017-01112-9} {\bibfield  {journal}
		{\bibinfo  {journal} {Sci. Rep.}\ }\textbf {\bibinfo {volume} {7}},\ \bibinfo
		{pages} {1306} (\bibinfo {year} {2017})}\BibitemShut {NoStop}%
	\bibitem [{\citenamefont {Qu}\ \emph {et~al.}(2015)\citenamefont {Qu},
		\citenamefont {Huang},\ and\ \citenamefont {Chien}}]{qu2015}%
	\BibitemOpen
	\bibfield  {author} {\bibinfo {author} {\bibfnamefont {D.}~\bibnamefont
			{Qu}}, \bibinfo {author} {\bibfnamefont {S.~Y.}\ \bibnamefont {Huang}}, \
		and\ \bibinfo {author} {\bibfnamefont {C.~L.}\ \bibnamefont {Chien}},\ }\href
	{\doibase 10.1103/PhysRevB.92.020418} {\bibfield  {journal} {\bibinfo
			{journal} {Phys. Rev. B}\ }\textbf {\bibinfo {volume} {92}},\ \bibinfo
		{pages} {020418} (\bibinfo {year} {2015})}\BibitemShut {NoStop}%
	\bibitem [{\citenamefont {Du}\ \emph {et~al.}(2014)\citenamefont {Du},
		\citenamefont {Wang}, \citenamefont {Yang},\ and\ \citenamefont
		{Hammel}}]{du2014}%
	\BibitemOpen
	\bibfield  {author} {\bibinfo {author} {\bibfnamefont {C.}~\bibnamefont
			{Du}}, \bibinfo {author} {\bibfnamefont {H.}~\bibnamefont {Wang}}, \bibinfo
		{author} {\bibfnamefont {F.}~\bibnamefont {Yang}}, \ and\ \bibinfo {author}
		{\bibfnamefont {P.~C.}\ \bibnamefont {Hammel}},\ }\href {\doibase
		10.1103/PhysRevB.90.140407} {\bibfield  {journal} {\bibinfo  {journal} {Phys.
				Rev. B}\ }\textbf {\bibinfo {volume} {90}},\ \bibinfo {pages} {140407}
		(\bibinfo {year} {2014})}\BibitemShut {NoStop}%
	\bibitem [{\citenamefont {Miao}\ \emph {et~al.}(2013)\citenamefont {Miao},
		\citenamefont {Huang}, \citenamefont {Qu},\ and\ \citenamefont
		{Chien}}]{miao2013}%
	\BibitemOpen
	\bibfield  {author} {\bibinfo {author} {\bibfnamefont {B.~F.}\ \bibnamefont
			{Miao}}, \bibinfo {author} {\bibfnamefont {S.~Y.}\ \bibnamefont {Huang}},
		\bibinfo {author} {\bibfnamefont {D.}~\bibnamefont {Qu}}, \ and\ \bibinfo
		{author} {\bibfnamefont {C.~L.}\ \bibnamefont {Chien}},\ }\href {\doibase
		10.1103/PhysRevLett.111.066602} {\bibfield  {journal} {\bibinfo  {journal}
			{Phys. Rev. Lett.}\ }\textbf {\bibinfo {volume} {111}},\ \bibinfo {pages}
		{066602} (\bibinfo {year} {2013})}\BibitemShut {NoStop}%
	\bibitem [{\citenamefont {An}\ \emph {et~al.}(2016)\citenamefont {An},
		\citenamefont {Kageyama}, \citenamefont {Kanno}, \citenamefont {Enishi},\
		and\ \citenamefont {Ando}}]{an2016}%
	\BibitemOpen
	\bibfield  {author} {\bibinfo {author} {\bibfnamefont {H.}~\bibnamefont
			{An}}, \bibinfo {author} {\bibfnamefont {Y.}~\bibnamefont {Kageyama}},
		\bibinfo {author} {\bibfnamefont {Y.}~\bibnamefont {Kanno}}, \bibinfo
		{author} {\bibfnamefont {N.}~\bibnamefont {Enishi}}, \ and\ \bibinfo {author}
		{\bibfnamefont {K.}~\bibnamefont {Ando}},\ }\href@noop {} {\bibfield
		{journal} {\bibinfo  {journal} {Nat. Commun.}\ }\textbf {\bibinfo {volume}
			{7}},\ \bibinfo {pages} {13069} (\bibinfo {year} {2016})}\BibitemShut
	{NoStop}%
	\bibitem [{\citenamefont {Gao}\ \emph {et~al.}(2018)\citenamefont {Gao},
		\citenamefont {Qaiumzadeh}, \citenamefont {An}, \citenamefont {Musha},
		\citenamefont {Kageyama}, \citenamefont {Shi},\ and\ \citenamefont
		{Ando}}]{gao2018}%
	\BibitemOpen
	\bibfield  {author} {\bibinfo {author} {\bibfnamefont {T.}~\bibnamefont
			{Gao}}, \bibinfo {author} {\bibfnamefont {A.}~\bibnamefont {Qaiumzadeh}},
		\bibinfo {author} {\bibfnamefont {H.}~\bibnamefont {An}}, \bibinfo {author}
		{\bibfnamefont {A.}~\bibnamefont {Musha}}, \bibinfo {author} {\bibfnamefont
			{Y.}~\bibnamefont {Kageyama}}, \bibinfo {author} {\bibfnamefont
			{J.}~\bibnamefont {Shi}}, \ and\ \bibinfo {author} {\bibfnamefont
			{K.}~\bibnamefont {Ando}},\ }\href {\doibase 10.1103/PhysRevLett.121.017202}
	{\bibfield  {journal} {\bibinfo  {journal} {Phys. Rev. Lett.}\ }\textbf
		{\bibinfo {volume} {121}},\ \bibinfo {pages} {017202} (\bibinfo {year}
		{2018})}\BibitemShut {NoStop}%
	\bibitem [{\citenamefont {Slater}\ and\ \citenamefont
		{Koster}(1954)}]{slater1954}%
	\BibitemOpen
	\bibfield  {author} {\bibinfo {author} {\bibfnamefont {J.~C.}\ \bibnamefont
			{Slater}}\ and\ \bibinfo {author} {\bibfnamefont {G.~F.}\ \bibnamefont
			{Koster}},\ }\href {\doibase 10.1103/PhysRev.94.1498} {\bibfield  {journal}
		{\bibinfo  {journal} {Phys. Rev.}\ }\textbf {\bibinfo {volume} {94}},\
		\bibinfo {pages} {1498} (\bibinfo {year} {1954})}\BibitemShut {NoStop}%
	\bibitem [{\citenamefont {Papaconstantopoulos}(2015)}]{handbook}%
	\BibitemOpen
	\bibfield  {author} {\bibinfo {author} {\bibfnamefont {D.~A.}\ \bibnamefont
			{Papaconstantopoulos}},\ }in\ \href@noop {} {\emph {\bibinfo {booktitle}
			{Handbook of the band structure of elemental solids: from Z = 1 to Z = 112:
				2nd ed.}}}\ (\bibinfo  {publisher} {Springer},\ \bibinfo {year}
	{2015})\BibitemShut {NoStop}%
	\bibitem [{\citenamefont {{Bihlmayer}}\ \emph {et~al.}(2006)\citenamefont
		{{Bihlmayer}}, \citenamefont {{Koroteev}}, \citenamefont {{Echenique}},
		\citenamefont {{Chulkov}},\ and\ \citenamefont {{Bl{\"u}gel}}}]{bihlmayer}%
	\BibitemOpen
	\bibfield  {author} {\bibinfo {author} {\bibfnamefont {G.}~\bibnamefont
			{{Bihlmayer}}}, \bibinfo {author} {\bibfnamefont {Y.~M.}\ \bibnamefont
			{{Koroteev}}}, \bibinfo {author} {\bibfnamefont {P.~M.}\ \bibnamefont
			{{Echenique}}}, \bibinfo {author} {\bibfnamefont {E.~V.}\ \bibnamefont
			{{Chulkov}}}, \ and\ \bibinfo {author} {\bibfnamefont {S.}~\bibnamefont
			{{Bl{\"u}gel}}},\ }\href {\doibase 10.1016/j.susc.2006.01.098} {\bibfield
		{journal} {\bibinfo  {journal} {Surf. Sci.}\ }\textbf {\bibinfo {volume}
			{600}},\ \bibinfo {pages} {3888} (\bibinfo {year} {2006})}\BibitemShut
	{NoStop}%
	\bibitem [{\citenamefont {Stamm}\ \emph {et~al.}(2017)\citenamefont {Stamm},
		\citenamefont {Murer}, \citenamefont {Berritta}, \citenamefont {Feng},
		\citenamefont {Gabureac}, \citenamefont {Oppeneer},\ and\ \citenamefont
		{Gambardella}}]{stamm2017}%
	\BibitemOpen
	\bibfield  {author} {\bibinfo {author} {\bibfnamefont {C.}~\bibnamefont
			{Stamm}}, \bibinfo {author} {\bibfnamefont {C.}~\bibnamefont {Murer}},
		\bibinfo {author} {\bibfnamefont {M.}~\bibnamefont {Berritta}}, \bibinfo
		{author} {\bibfnamefont {J.}~\bibnamefont {Feng}}, \bibinfo {author}
		{\bibfnamefont {M.}~\bibnamefont {Gabureac}}, \bibinfo {author}
		{\bibfnamefont {P.~M.}\ \bibnamefont {Oppeneer}}, \ and\ \bibinfo {author}
		{\bibfnamefont {P.}~\bibnamefont {Gambardella}},\ }\href {\doibase
		10.1103/PhysRevLett.119.087203} {\bibfield  {journal} {\bibinfo  {journal}
			{Phys. Rev. Lett.}\ }\textbf {\bibinfo {volume} {119}},\ \bibinfo {pages}
		{087203} (\bibinfo {year} {2017})}\BibitemShut {NoStop}%
	\bibitem [{not()}]{note}%
	\BibitemOpen
	\href@noop {} {}\bibinfo {note} {We remark that although detailed behavior
		may be different depending on the choice of the basis, our general conclusion
		that the OHE and SHE scales with the orbital texture strength is unchanged.
		See Appendix for details.}\BibitemShut {Stop}%
	\bibitem [{\citenamefont {Ashcroft}\ and\ \citenamefont
		{Mermin}(1976)}]{ashcroft}%
	\BibitemOpen
	\bibfield  {author} {\bibinfo {author} {\bibfnamefont {N.}~\bibnamefont
			{Ashcroft}}\ and\ \bibinfo {author} {\bibfnamefont {N.}~\bibnamefont
			{Mermin}},\ }\href@noop {} {\emph {\bibinfo {title} {Solid State Physics}}}\
	(\bibinfo  {publisher} {Holt, Rinehart and Winston},\ \bibinfo {year}
	{1976})\BibitemShut {NoStop}%
	\bibitem [{\citenamefont {Go}\ \emph {et~al.}(2017)\citenamefont {Go},
		\citenamefont {Hanke}, \citenamefont {Buhl}, \citenamefont {Freimuth},
		\citenamefont {Bihlmayer}, \citenamefont {Lee}, \citenamefont {Mokrousov},\
		and\ \citenamefont {Bl{\"{u}}gel}}]{go2017}%
	\BibitemOpen
	\bibfield  {author} {\bibinfo {author} {\bibfnamefont {D.}~\bibnamefont
			{Go}}, \bibinfo {author} {\bibfnamefont {J.~P.}\ \bibnamefont {Hanke}},
		\bibinfo {author} {\bibfnamefont {P.~M.}\ \bibnamefont {Buhl}}, \bibinfo
		{author} {\bibfnamefont {F.}~\bibnamefont {Freimuth}}, \bibinfo {author}
		{\bibfnamefont {G.}~\bibnamefont {Bihlmayer}}, \bibinfo {author}
		{\bibfnamefont {H.~W.}\ \bibnamefont {Lee}}, \bibinfo {author} {\bibfnamefont
			{Y.}~\bibnamefont {Mokrousov}}, \ and\ \bibinfo {author} {\bibfnamefont
			{S.}~\bibnamefont {Bl{\"{u}}gel}},\ }\href@noop {} {\bibfield  {journal}
		{\bibinfo  {journal} {Sci. Rep.}\ }\textbf {\bibinfo {volume} {7}},\ \bibinfo
		{pages} {1} (\bibinfo {year} {2017})}\BibitemShut {NoStop}%
	\bibitem [{\citenamefont {Wimmer}\ \emph {et~al.}(1981)\citenamefont {Wimmer},
		\citenamefont {Krakauer}, \citenamefont {Weinert},\ and\ \citenamefont
		{Freeman}}]{wimmer1981}%
	\BibitemOpen
	\bibfield  {author} {\bibinfo {author} {\bibfnamefont {E.}~\bibnamefont
			{Wimmer}}, \bibinfo {author} {\bibfnamefont {H.}~\bibnamefont {Krakauer}},
		\bibinfo {author} {\bibfnamefont {M.}~\bibnamefont {Weinert}}, \ and\
		\bibinfo {author} {\bibfnamefont {A.~J.}\ \bibnamefont {Freeman}},\ }\href
	{\doibase 10.1103/PhysRevB.24.864} {\bibfield  {journal} {\bibinfo  {journal}
			{Phys. Rev. B}\ }\textbf {\bibinfo {volume} {24}},\ \bibinfo {pages} {864}
		(\bibinfo {year} {1981})}\BibitemShut {NoStop}%
	\bibitem [{\citenamefont {Hanke}\ \emph {et~al.}(2016)\citenamefont {Hanke},
		\citenamefont {Freimuth}, \citenamefont {Nandy}, \citenamefont {Zhang},
		\citenamefont {Bl\"ugel},\ and\ \citenamefont {Mokrousov}}]{hanke2016}%
	\BibitemOpen
	\bibfield  {author} {\bibinfo {author} {\bibfnamefont {J.-P.}\ \bibnamefont
			{Hanke}}, \bibinfo {author} {\bibfnamefont {F.}~\bibnamefont {Freimuth}},
		\bibinfo {author} {\bibfnamefont {A.~K.}\ \bibnamefont {Nandy}}, \bibinfo
		{author} {\bibfnamefont {H.}~\bibnamefont {Zhang}}, \bibinfo {author}
		{\bibfnamefont {S.}~\bibnamefont {Bl\"ugel}}, \ and\ \bibinfo {author}
		{\bibfnamefont {Y.}~\bibnamefont {Mokrousov}},\ }\href@noop {} {\bibfield
		{journal} {\bibinfo  {journal} {Phys. Rev. B}\ }\textbf {\bibinfo {volume}
			{94}},\ \bibinfo {pages} {121114} (\bibinfo {year} {2016})}\BibitemShut
	{NoStop}%
	\bibitem [{\citenamefont {O'Brien}\ and\ \citenamefont
		{Tonner}(1994)}]{obrien1994}%
	\BibitemOpen
	\bibfield  {author} {\bibinfo {author} {\bibfnamefont {W.~L.}\ \bibnamefont
			{O'Brien}}\ and\ \bibinfo {author} {\bibfnamefont {B.~P.}\ \bibnamefont
			{Tonner}},\ }\href {\doibase 10.1103/PhysRevB.50.12672} {\bibfield  {journal}
		{\bibinfo  {journal} {Phys. Rev. B}\ }\textbf {\bibinfo {volume} {50}},\
		\bibinfo {pages} {12672} (\bibinfo {year} {1994})}\BibitemShut {NoStop}%
	\bibitem [{\citenamefont {Bonetti}(2017)}]{bonetti2017}%
	\BibitemOpen
	\bibfield  {author} {\bibinfo {author} {\bibfnamefont {S.}~\bibnamefont
			{Bonetti}},\ }\href {http://stacks.iop.org/0953-8984/29/i=13/a=133004}
	{\bibfield  {journal} {\bibinfo  {journal} {J. Phys. Condensed Matter}\
		}\textbf {\bibinfo {volume} {29}},\ \bibinfo {pages} {133004} (\bibinfo
		{year} {2017})}\BibitemShut {NoStop}%
	\bibitem [{\citenamefont {Kato}\ \emph {et~al.}(2004)\citenamefont {Kato},
		\citenamefont {Myers}, \citenamefont {Gossard},\ and\ \citenamefont
		{Awschalom}}]{kato2004}%
	\BibitemOpen
	\bibfield  {author} {\bibinfo {author} {\bibfnamefont {Y.~K.}\ \bibnamefont
			{Kato}}, \bibinfo {author} {\bibfnamefont {R.~C.}\ \bibnamefont {Myers}},
		\bibinfo {author} {\bibfnamefont {A.~C.}\ \bibnamefont {Gossard}}, \ and\
		\bibinfo {author} {\bibfnamefont {D.~D.}\ \bibnamefont {Awschalom}},\ }\href
	{\doibase 10.1126/science.1105514} {\bibfield  {journal} {\bibinfo  {journal}
			{Science}\ }\textbf {\bibinfo {volume} {306}},\ \bibinfo {pages} {1910}
		(\bibinfo {year} {2004})}\BibitemShut {NoStop}%
	\bibitem [{\citenamefont {Puebla}\ \emph {et~al.}(2017)\citenamefont {Puebla},
		\citenamefont {Auvray}, \citenamefont {Xu}, \citenamefont {Rana},
		\citenamefont {Albouy}, \citenamefont {Tsai}, \citenamefont {Kondou},
		\citenamefont {Tatara},\ and\ \citenamefont {Otani}}]{puebla2017}%
	\BibitemOpen
	\bibfield  {author} {\bibinfo {author} {\bibfnamefont {J.}~\bibnamefont
			{Puebla}}, \bibinfo {author} {\bibfnamefont {F.}~\bibnamefont {Auvray}},
		\bibinfo {author} {\bibfnamefont {M.}~\bibnamefont {Xu}}, \bibinfo {author}
		{\bibfnamefont {B.}~\bibnamefont {Rana}}, \bibinfo {author} {\bibfnamefont
			{A.}~\bibnamefont {Albouy}}, \bibinfo {author} {\bibfnamefont
			{H.}~\bibnamefont {Tsai}}, \bibinfo {author} {\bibfnamefont {K.}~\bibnamefont
			{Kondou}}, \bibinfo {author} {\bibfnamefont {G.}~\bibnamefont {Tatara}}, \
		and\ \bibinfo {author} {\bibfnamefont {Y.}~\bibnamefont {Otani}},\ }\href
	{\doibase 10.1063/1.4990113} {\bibfield  {journal} {\bibinfo  {journal}
			{Appl. Phys. Lett.}\ }\textbf {\bibinfo {volume} {111}},\ \bibinfo {pages}
		{092402} (\bibinfo {year} {2017})}\BibitemShut {NoStop}%
	\bibitem [{\citenamefont {Liu}\ \emph {et~al.}(2018)\citenamefont {Liu},
		\citenamefont {Besbas}, \citenamefont {Wang}, \citenamefont {He},
		\citenamefont {Chen}, \citenamefont {Zhu}, \citenamefont {Wu}, \citenamefont
		{Lee}, \citenamefont {Wang}, \citenamefont {Moon}, \citenamefont {Koirala},
		\citenamefont {Oh},\ and\ \citenamefont {Yang}}]{liu2018}%
	\BibitemOpen
	\bibfield  {author} {\bibinfo {author} {\bibfnamefont {Y.}~\bibnamefont
			{Liu}}, \bibinfo {author} {\bibfnamefont {J.}~\bibnamefont {Besbas}},
		\bibinfo {author} {\bibfnamefont {Y.}~\bibnamefont {Wang}}, \bibinfo {author}
		{\bibfnamefont {P.}~\bibnamefont {He}}, \bibinfo {author} {\bibfnamefont
			{M.}~\bibnamefont {Chen}}, \bibinfo {author} {\bibfnamefont {D.}~\bibnamefont
			{Zhu}}, \bibinfo {author} {\bibfnamefont {Y.}~\bibnamefont {Wu}}, \bibinfo
		{author} {\bibfnamefont {J.~M.}\ \bibnamefont {Lee}}, \bibinfo {author}
		{\bibfnamefont {L.}~\bibnamefont {Wang}}, \bibinfo {author} {\bibfnamefont
			{J.}~\bibnamefont {Moon}}, \bibinfo {author} {\bibfnamefont {N.}~\bibnamefont
			{Koirala}}, \bibinfo {author} {\bibfnamefont {S.}~\bibnamefont {Oh}}, \ and\
		\bibinfo {author} {\bibfnamefont {H.}~\bibnamefont {Yang}},\ }\href@noop {}
	{\bibfield  {journal} {\bibinfo  {journal} {Nat. Commun.}\ }\textbf {\bibinfo
			{volume} {9}} (\bibinfo {year} {2018})}\BibitemShut {NoStop}%
	\bibitem [{\citenamefont {Bernevig}\ \emph {et~al.}(2005)\citenamefont
		{Bernevig}, \citenamefont {Hughes},\ and\ \citenamefont
		{Zhang}}]{bernevig2005b}%
	\BibitemOpen
	\bibfield  {author} {\bibinfo {author} {\bibfnamefont {B.~A.}\ \bibnamefont
			{Bernevig}}, \bibinfo {author} {\bibfnamefont {T.~L.}\ \bibnamefont
			{Hughes}}, \ and\ \bibinfo {author} {\bibfnamefont {S.-C.}\ \bibnamefont
			{Zhang}},\ }\href {\doibase 10.1103/PhysRevLett.95.066601} {\bibfield
		{journal} {\bibinfo  {journal} {Phys. Rev. Lett.}\ }\textbf {\bibinfo
			{volume} {95}},\ \bibinfo {pages} {066601} (\bibinfo {year}
		{2005})}\BibitemShut {NoStop}%
	\bibitem [{\citenamefont {Kim}\ \emph {et~al.}(2008)\citenamefont {Kim},
		\citenamefont {Jin}, \citenamefont {Moon}, \citenamefont {Kim}, \citenamefont
		{Park}, \citenamefont {Leem}, \citenamefont {Yu}, \citenamefont {Noh},
		\citenamefont {Kim}, \citenamefont {Oh}, \citenamefont {Park}, \citenamefont
		{Durairaj}, \citenamefont {Cao},\ and\ \citenamefont {Rotenberg}}]{kim2008}%
	\BibitemOpen
	\bibfield  {author} {\bibinfo {author} {\bibfnamefont {B.~J.}\ \bibnamefont
			{Kim}}, \bibinfo {author} {\bibfnamefont {H.}~\bibnamefont {Jin}}, \bibinfo
		{author} {\bibfnamefont {S.~J.}\ \bibnamefont {Moon}}, \bibinfo {author}
		{\bibfnamefont {J.-Y.}\ \bibnamefont {Kim}}, \bibinfo {author} {\bibfnamefont
			{B.-G.}\ \bibnamefont {Park}}, \bibinfo {author} {\bibfnamefont {C.~S.}\
			\bibnamefont {Leem}}, \bibinfo {author} {\bibfnamefont {J.}~\bibnamefont
			{Yu}}, \bibinfo {author} {\bibfnamefont {T.~W.}\ \bibnamefont {Noh}},
		\bibinfo {author} {\bibfnamefont {C.}~\bibnamefont {Kim}}, \bibinfo {author}
		{\bibfnamefont {S.-J.}\ \bibnamefont {Oh}}, \bibinfo {author} {\bibfnamefont
			{J.-H.}\ \bibnamefont {Park}}, \bibinfo {author} {\bibfnamefont
			{V.}~\bibnamefont {Durairaj}}, \bibinfo {author} {\bibfnamefont
			{G.}~\bibnamefont {Cao}}, \ and\ \bibinfo {author} {\bibfnamefont
			{E.}~\bibnamefont {Rotenberg}},\ }\href {\doibase
		10.1103/PhysRevLett.101.076402} {\bibfield  {journal} {\bibinfo  {journal}
			{Phys. Rev. Lett.}\ }\textbf {\bibinfo {volume} {101}},\ \bibinfo {pages}
		{076402} (\bibinfo {year} {2008})}\BibitemShut {NoStop}%
	\bibitem [{fle()}]{fleur}%
	\BibitemOpen
	\href@noop {} {}\bibinfo {howpublished}
	{\url{http://www.flapw.de}}\BibitemShut {NoStop}%
	\bibitem [{\citenamefont {Perdew}\ \emph {et~al.}(1996)\citenamefont {Perdew},
		\citenamefont {Burke},\ and\ \citenamefont {Ernzerhof}}]{perdew1996}%
	\BibitemOpen
	\bibfield  {author} {\bibinfo {author} {\bibfnamefont {J.~P.}\ \bibnamefont
			{Perdew}}, \bibinfo {author} {\bibfnamefont {K.}~\bibnamefont {Burke}}, \
		and\ \bibinfo {author} {\bibfnamefont {M.}~\bibnamefont {Ernzerhof}},\ }\href
	{\doibase 10.1103/PhysRevLett.77.3865} {\bibfield  {journal} {\bibinfo
			{journal} {Phys. Rev. Lett.}\ }\textbf {\bibinfo {volume} {77}},\ \bibinfo
		{pages} {3865} (\bibinfo {year} {1996})}\BibitemShut {NoStop}%
	\bibitem [{\citenamefont {Li}\ \emph {et~al.}(1990)\citenamefont {Li},
		\citenamefont {Freeman}, \citenamefont {Jansen},\ and\ \citenamefont
		{Fu}}]{li1990}%
	\BibitemOpen
	\bibfield  {author} {\bibinfo {author} {\bibfnamefont {C.}~\bibnamefont
			{Li}}, \bibinfo {author} {\bibfnamefont {A.~J.}\ \bibnamefont {Freeman}},
		\bibinfo {author} {\bibfnamefont {H.~J.~F.}\ \bibnamefont {Jansen}}, \ and\
		\bibinfo {author} {\bibfnamefont {C.~L.}\ \bibnamefont {Fu}},\ }\href
	{\doibase 10.1103/PhysRevB.42.5433} {\bibfield  {journal} {\bibinfo
			{journal} {Phys. Rev. B}\ }\textbf {\bibinfo {volume} {42}},\ \bibinfo
		{pages} {5433} (\bibinfo {year} {1990})}\BibitemShut {NoStop}%
\end{thebibliography}

%

\end{document}